\documentclass[showpacs,amsmath,amssymb]{revtex4}
\usepackage{graphicx}
\begin{document}

\title{Simultaneous description of four positive and four negative parity bands}

\author{A. A. Raduta$^{a), b)}$, Al. H. Raduta $^{b)}$ and C. M. Raduta $^{b)}$ }

\address{$^{a)}$Department of Theoretical Physics and Mathematics,
Bucharest University, POBox MG11, Romania}

\address{$^{b)}$Department of Theoretical Physics,
Institute of Physics and Nuclear Engineering, Bucharest, POBox MG6,
 Romania}

\date{\today}

\begin{abstract} The extended coherent state model is further extended
in order to describe  two dipole bands of different parities.
The formalism provides a consistent description of eight rotational
bands. A unified description for spherical,
transitional and deformed nuclei is possible.
Projecting out the angular momentum and parity from a sole state,
the $K^{\pi}=1^+$ band acquires a magnetic character, while
the electric properties prevail for the other band. Signatures for a
static octupole deformation in some states of the dipole bands  are
pointed out. Some properties which distinguish between the dipole band states
and states
of the same parity but belonging to other bands are mentioned.
Interesting features
concerning the decay properties of the two bands are found. Numerical applications are made
for $^{158}$Gd, $^{172}$Yb, $^{228,232}$Th, $^{226}$Ra, $^{238}$U and $^{238}$Pu, and the results
are compared with the available data.

\end{abstract}



\maketitle
\section{Introduction}
\renewcommand{\theequation}{1.\arabic{equation}}
\setcounter{equation}{0}

\label{sec:level1}

The field of negative parity bands became very attractive when
the first suggestions for a static octupole deformation were advanced by
Chassman\cite{Cha}, and Moler and Nix \cite{Mol}.
Since a nuclear shape with octupole deformation does not exhibit a space
reflection
symmetry and, on the other hand, a spontaneously broken symmetry leads to a new
nuclear phase,
one expects that the octupole deformed nuclei have specific properties.
The main achievements of this field have been reviewed in Refs. \cite{Roho,ButNaz,Frau}.

Identifying the nuclei which have static octupole
deformation seems to be a difficult task.
Indeed, because there is no measurable quantity for the octupole deformation,
some indirect information about this variable should be found.
Several properties  are considered as signatures for octupole 
deformation:
a) In some nuclei like $^{218}$Ra, the state $1^-$, the head of the
$K^{\pi}=0^-$ band, has a very low position, and this is an
indication that the potential energy has a flat minimum, as a
function of the octupole deformation.
b) The parity alternating structure in ground and the lowest $0^-$ bands
suggests that 
the two bands may be viewed as being projected from a sole deformed intrinsic
state, 
exhibiting both quadrupole and octupole deformations. 
c)  A nuclear surface with quadrupole and octupole deformations might have the
centre of charge in a different position than the centre of mass, which results
 in having
an electric dipole moment which may excite the state $1^-$ from the ground state, 
with a large probability.
The list is not complete and thereby any
new signature for this new nuclear phase deserves a special attention.

Few years ago we considered this subject within a phenomenological
framework. 
Thus, in Refs. \cite{Rad8,Rad9,Rad10,Rad110,Rad13} we extended the coherent state
model (CSM) \cite{Rad11,Rad12} to the negative parity bands. To the lowest positive
parity bands, named ground ($g^{+}$), beta ($\beta^{+}$) and  gamma ($\gamma^{+}$),
one associates
three negative bands, $g^{-}$, $\beta^{-}$, $\gamma^{-}$, respectively.
The six bands are obtained by projecting out the angular momentum and the
parity from three orthogonal functions which exhibit both quadrupole and
octupole deformations. An effective boson Hamiltonian is considered in the
space of angular momentum and parity projected states.
The phenomenological boson model called Extended Coherent State Model (ECSM), has been successfully applied
to a large number of nuclei, some of them being suspected to exhibit a static octupole deformation while some of them
having vibrational octupole bands. Some signatures for a static octupole deformation in
the excited bands have been pointed out.

In the present paper we extend even more the coherent state model by adding a new pair of parity partner bands.
These are characterised by $K^{\pi}=1^+$ and $K^{\pi}=1^-$. Also two new terms
are added to the model Hamiltonian without altering its
 effective character, whose strength are fixed by fitting some particular
 data for the new bands.

The new extension is presented according to the following plan.
In Section II, a brief description of the CSM and ECSM is given. The scope consists in having a self-standing
work and on the other hand in collecting the necessary definitions and notations.
In section III, the ingredients of the new extension are presented in extenso,
i.e. the properties of the states which enlarge the model boson space as well as the
corrective terms of the model Hamiltonian and their matrix elements are analytically given.
In Section IV, we discuss the numerical application for seven nuclei. Since for
$^{172}$Yb and $^{226}$Ra, some results were reported in two earlier
publications,
here we consider only the new results.
A summary of the results and the  final conclusions are presented in Section V.

\section{Brief review of the coherent state model and its extended version}
\renewcommand{\theequation}{2.\arabic{equation}}
\setcounter{equation}{0}

\label{sec:level2}

\subsection{The coherent state model (CSM)}
In the beginning of eighties, one of the present authors (A. A. R) proposed,
in collaboration, a
phenomenological model to describe the main properties of the first three
collective bands i.e., ground, beta and gamma bands \cite{Rad11,Rad12}.
The model space was generated through a projection procedure from three orthogonal deformed states. The choice was made so that several criteria required by the existent data are fulfilled.
The states are built up with quadrupole bosons and therefore we are dealing
with those properties which are determined by the collective motion of the quadrupole degrees of freedom.

We suppose that the intrinsic ground state is described by a coherent state
of Glauber type corresponding to the zeroth component of the quadrupole boson
operator $b_{2\mu}$. The other two generating functions are the simplest polynomial excitations of the intrinsic ground state, chosen in such a way that the orthogonality condition is satisfied before and after projection.
To each intrinsic state one associates an infinite rotational band.
In two of these bands the spin sequence is $0^+, 2^+, 4^+, 6^+,..$ etc.,
and therefore they correspond to the ground (the lowest one) and to the beta
bands, respectively. The third one
involves all angular momenta larger or equal to 2, and is describing, in the first order of approximation, the gamma band.
The intrinsic states depend on a real parameter d which simulates the nuclear deformation. In the spherical limit, i.e. d goes to zero, the projected states are multi-phonon states of highest, second and
third highest seniority, respectively. In the large deformation regime,
conventionally called rotational limit
(d equal to 3 means already a rotational limit), the model states behave like a Wigner function, which fully agrees the behaviour prescribed by the liquid drop model. The correspondence between the states in the spherical and rotational limits is achieved by a smooth variation of the deformation parameter. This correspondence agrees perfectly with the semi-empirical rule of Sheline\cite{She}
 and Sakai\cite{Saka}
, concerning the linkage of the ground, beta and gamma band states and the member of multi-phonon states from the vibrational limit. This property is very important 
when one wants to describe the gross features of the reduced probabilities
for the intra and inter bands transitions.

In this restricted collective model space an effective boson Hamiltonian is constructed. A very simple Hamiltonian was found, which has only one off-diagonal matrix element, namely that one connecting the states from the ground and the gamma bands. 
\begin{eqnarray}
H_{CSM}&=&H_2^{\prime}+\lambda\hat{J}^2_2,\nonumber\\
H_2'&=&A_1(22\hat{N}_2+5\Omega^{\dagger}_{\beta'}\Omega_{\beta'})+A_2\Omega^{\dagger}_{\beta}\Omega_{\beta},
\end{eqnarray}
where $\hat{N}_2$ denotes the quadrupole boson number operator
\begin{equation}
\hat{N}_2=\sum_{-2\leq m\leq 2}b_{2m}^{\dagger}b_{2m},
\end{equation}
while $\Omega^{\dagger}_{\beta'}$ and  $\Omega^{\dagger}_{\beta}$ stand for the following second and third degree scalar polynomials:
 \begin{eqnarray}
  \Omega^{\dagger}_{\beta'}&=&(b^{\dagger}_2 b^{\dagger}_2)_0-\frac{d^2}{\sqrt{5}},\nonumber\\
  \Omega^{\dagger}_{\beta}&=&(b^{\dagger}_2 b^{\dagger}_2 b^{\dagger}_2)_0+\frac{3d}{\sqrt{14}}(b^{\dagger}_2 b^{\dagger}_2)_0-\frac{d^3}{\sqrt{70}}.
\label{Omega}
  \end{eqnarray}
The angular momentum carried by the quadrupole bosons, is denoted by $\hat{J}_2$.
The boson states space is spanned by the projected states:
\begin{equation}
\varphi^{(i)}_{JM}=N^{(i)}_JP^J_{MK}\psi_{i},~~i=g,\beta,\gamma,
\end{equation}
where the intrinsic states are:
\begin{equation}
\psi_g=e^{d(b^{\dag}_{20}-b_{20})}|0\rangle,~~
\psi_{\beta}=\Omega^{\dag}_{\beta}\psi_g,~~
\psi_{\gamma}=\Omega^{\dag}_{\gamma}\psi_g.
\end{equation}
The excitation operator $\Omega^{\dag}_{\beta}$ is given by Eq.(\ref{Omega}),
while
the operator $\Omega^{\dag}_{\gamma}$, which excites the gamma band states, is:
\begin{equation}
\Omega^{\dag}_{\gamma}=(b^{\dag}_2b^{\dag}_2)_{22}+d\sqrt{\frac{2}{7}}b^{\dag}_{22}.
\end{equation}
The angular momentum projection operator is defined by:
\begin{equation}
P^J_{MK}=\frac{2J+1}{8\pi^2}\int D^{J^*}_{MK}(\Omega )\hat{R}(\Omega)d\Omega ,
\label{PJMK}
\end{equation}
where the standard notations for the Wigner function and the rotation operator 
corresponding to the Eulerian angles $\Omega$, have been used.

The eigenvalues of the effective Hamiltonian in the restricted space of projected states have been analytically studied in both spherical and rotational limit. Compact formulae for transition probabilities in the two extreme limits have been also derived. This model has been successfully applied for a large number of nuclei  from transitional and well deformed regions.
It is worth to mention that by varying the deformation parameter and the
parameters defining the effective Hamiltonian one can realistically describe
nuclei satisfying various symmetries like, SU(5) (Sm region)\cite{RadSand},
O(6) (Pt region)\cite{Rad11,Rad12}, SU3
(Th region)\cite{RaSab}, triaxial rotor (Ba, Xe isotopes)\cite{UliRad}.
This model has been extended by including the coupling to the individual
degrees of freedom \cite{RadCels}.
In this way the spectroscopic properties in the region of back-bending were quantitatively described.

The extension of the CSM formalism, which will be presented here,
considers a composite system of quadrupole and octupole bosons.
\subsection{The extended coherent state model (ECSM)}
 The CSM formalism was generalised by assuming that the intrinsic ground
 state exhibits not only a quadrupole deformation but also an octupole one.
 Since the other bands, beta and gamma, are excited from the ground state,
 they also have this property.
The octupole deformation is described by means of an axially symmetric
coherent state for the octupole bosons $b^{\dag}_{30}$. Thus, the intrinsic 
states for ground, beta and gamma bands are:
\begin{equation}
\Psi_g=e^{f(b^{\dag}_{30}-b_{30})}e^{d(b^{\dag}_{20}-b_{20})}|0\rangle_{(3)}|0\rangle_{(2)},~
\Psi_{\beta}=\Omega^{\dag}_{\beta}\Psi_g,~
\Psi_{\gamma}=\Omega^{\dag}_{\gamma}\Psi_g. 
\end{equation}
The notation $|0\rangle_{(k)}$ stands for the vacuum state of the $2^k$-pole boson operators. Note that any of these states is a mixture of positive and negative parity states. Therefore they don't have good reflection symmetry.
Due to this feature the new extension of the CSM formalism has to project out not only the angular momentum but also the parity.
The parity projection affects only the factor function depending on octupole bosons.
Useful simplifications are achieved when  this factor function is separately
 treated.
The parity projected states are defined by:
\begin{equation}
\Psi^{(k)}_{oc}=P^{(k)}e^{f(b^{\dag}_{30}-b_{30})}|0\rangle_{(3)},k=\pm,
\end{equation}
where $P^{(k)}$ denotes the parity projection operator which is defined by its property that acting on a state consisting of a  series of boson operators
acting on the octupole vacuum, it selects only components with even powers in bosons if $k=+$ and odd components for $k=-$.
From the parity projected states one projects out, further, the components with good angular momentum:
\begin{equation}
\Psi^{(k)}_{oc; J_3M_3}=N^{(k)}_{oc; J_3}P^J_{M_30}\Psi^{(k)}_{oc}.
\end{equation} 
The factor $N^{(k)}_{oc;J_3}$ assures that the projected state has the norm equal
to unity. Its expression is given in Appendix A.

Then, the intrinsic states of good parity are defined by:
\begin{equation}
\Psi^{(k)}_i=\Psi^{(k)}_{oc}\Psi_i,~i=g,\beta,\gamma,~k=\pm.
\end{equation}
The member states of ground beta and gamma bands are projected from the corresponding intrinsic states:
\begin{eqnarray}
\varphi_{JM}^{(i,k)}&=&{\cal N}_J^{(i,k)}P_{MK_i}^J\Psi_i^{(k)},~~
K_i=2\delta_{i,\gamma},~~
k=\pm; i=g, \beta, \gamma,
\nonumber\\
&&J=(\delta_{i,g}+\delta_{i,\beta})(even \delta_{k,+}+ odd \delta_{k,-})
+\delta_{i,\gamma}(J\ge 2).
\end{eqnarray}
It can be shown that these projected states can be expressed by means of the
octupole factor projected states and the projected states characterising the
CSM formalism.
\begin{equation}
\varphi^{(i,k)}_{JM}={\cal N}^{(i,k)}_J\sum_{J_2,J_3}\left(N^{(k)}_{oc;J_3}
N^{(i)}_{J_2}\right)^{-1}C^{J_3~J_2~J}_{0~~K_i~~K_i}\left[\Psi^{(k)}_{oc;J_3}
\varphi^{(i)}_{J_2}\right]_{JM}, K_i=2\delta_{i,\gamma},~~
k=\pm; i=g, \beta, \gamma,
\end{equation}
The normalisation factor has the expression:
\begin{equation}
({\cal N}_J^{(i,k)})^{-2}=\sum_{J_2,J_3}(N_{oc;J_3}^{(k)}N_{J_2}^{(i)})^{-2}
(C_{0\hskip0.25cmK_i\hskip0.2cmK_i}^{J_3\hskip0.1cm J_2\hskip0.2cm J})^2,~
K_i=2\delta_{i,\gamma},~
k=\pm; i=g, \beta, \gamma.
\end{equation}
An effective boson Hamiltonian has been studied in the restricted collective
space generated by the six sets of projected states.
Note that from each of the three intrinsic states, one generates by projection
 two sets of states, one of positive and one of negative parity.
When the octupole deformation goes to zero, the resulting states are just those characterising the CSM model. In this limit we know already the effective quadrupole boson Hamiltonian. When the quadrupole deformation is going to zero
the system exhibits vibrations around an octupole deformed equilibrium shape.
We consider for the octupole Hamiltonian an harmonic structure since the
non-harmonic octupole terms can be simulated by the quadrupole anharmonicities.
 As for the  coupling between quadrupole and octupole bosons, we suppose that
 this can be described by a product between the octupole boson number operator,
 $\hat {N}_3$, and the quadrupole boson anharmonic terms which are involved in
 the CSM Hamiltonian. Indeed, it has been proved that including octupole anharmonicities in the coupling terms
 these terms provide an angular moment dependence for the corresponding matrix elements
 similar to the one already generated by the terms involving only the operator $\hat{N}_3$
 in the coupling with the quadrupole bosons.
 Also, the scalar product  of the angular momenta
 carried by the quadrupole ($\vec{J}_2$) and
 octupole bosons ($\vec{J}_3$), respectively,  and the total angular momentum
squared ( $\vec{J}^2$),  are included. Thus, the model Hamiltonian has the expression:
 
\begin{eqnarray}
H&=&H'_2+{\cal B}_1\hat{N}_3(22\hat{N}_2+5\Omega^{\dagger}_{\beta'}\Omega_{\beta'})+{\cal B}_2\hat{N}_3\Omega^{\dagger}_{\beta}\Omega_{\beta}
\nonumber\\
&&+{\cal B}_3\hat{N}_3+{\cal A}_{(J23)}\vec{J}_2\vec{J}_3+{\cal A}_J\vec{J}^2.
\end{eqnarray}
Detail arguments in favour of this choice are presented in our previous publication on this subject.
This Hamiltonian was used in Refs.\cite{Rad8,Rad9,Rad10} to study the ground
and $K^{\pi}=0^-$
bands. As was shown  in the quoted papers, the coupling term $\vec{J}_2\vec{J}_3$
is necessary in order to explain the low position of the state $1^-$ in
the
even-even Ra isotopes. Indeed, this term is attractive in the state $1^-$
while for other angular momenta  is repulsive.

Due to the specific structure of the CSM basis states the only
non-vanishing off-diagonal matrix elements are those connecting the
states of 
the ground and gamma and of the $0^-$ and $2^-$ bands.
The energies of the six bands are defined as eigenvalues of the model Hamiltonian in the model space of the
projected  states.
They  depend on the structure coefficients ${\cal A}_k$,(k=1,2,J,(J23)) and
${\cal B}_k$
,(k=1,3) defining the model Hamiltonian and the two deformation parameters,
d and f. 
Therefore there  are eight parameters which are to be determined, by  fitting the
data for excitation energies with the theoretical energies normalised to the
ground state energy.
For the  considered isotopes, the structure coefficients obtained in this manner
have a smooth behaviour when we change  A or Z.

The connection between the present description and the rotational bands, as defined in
the liquid drop model, was established in Refs. \cite{Rad12}.

Indeed, as shown in Ref.(\cite{Rad12}) the projected states are linear superposition of states with  definite K-quantum number.
Moreover, in the asymptotic limit of the deformation parameter a single K prevails
for each set of projected states,  associated to the intrinsic unprojected states respectively.
Assigning to each band that $K$ which labels the dominant component of the superposition quoted above,
one may assert that the projected states given by Eq.(2.13) comprises
two $K^{\pi}=0^+$, two $K^{\pi}=0^-$
one $K^{\pi}=2^+$ and one $K^{\pi}=2^-$ subsets. Note that the $K$ quantum
number is equal to the eigenvalue of $J_z$, corresponding to the unprojected states $\Psi_k$ with $k=g,\beta, \gamma$.
Thus, the symmetry breaking in the wave functions given by Eq.(2.8) is
equivalent to choosing an
auxiliary intrinsic frame of reference.

The bands associated to these quantum numbers are conventionally denoted by
$g^{\pm} (K^{\pi}=0^{\pm})$, $\beta^{\pm} (K^{\pi}=0^{\pm})$ and
 $\gamma^{\pm} (K^{\pi}=2^{\pm})$.
\section{The dipole bands}
\renewcommand{\theequation}{3.\arabic{equation}}
\setcounter{equation}{0}

\label{sec:level3}

Extending further the ECSM formalism, by adding some new bands with keeping
 the basic principles of CSM unaltered,
is a difficult task. Indeed, first one has to find an intrinsic
state which is orthogonal onto other three states defined so far. Moreover, the orthogonality property has to be preserved
also after projecting the angular momentum and parity.
Suppose that this step has been already overcome. The next step is then,
to extend the model
Hamiltonian by adding new terms which are mainly responsible for the description
of the new states.
The new Hamiltonian should be effective in the extended space of projected
states, i.e. the off diagonal matrix elements are either equal to zero or very small comparing them
with the diagonal ones.

In the present paper, we propose  the following solution for the
intrinsic state generating, through the angular momentum and parity projection,
the member states of the dipole bands:
\begin{equation}
\Psi^{(1,\pm)}=\Omega^{\dag}_3
b^{\dag}_{31} \Psi^{\mp}_{oc}\Psi_g , \rm{where}\;\;
\Omega^{\dag}_3=\left(b^{\dag}_3 b^{\dag}_3 \right)_0+\frac{f^2}{\sqrt{7}}.
\end{equation}
The states $\Psi^{(1,+)}$ and $\Psi^{(1,-)}$ are orthogonal since their scalar product involves
the overlap of components with different number of bosons.
Moreover, since $\Psi^{\pm}_{oc}$ are vacuum states for the operator $\Omega_3$,
the intrinsic states for the dipole bands are orthogonal onto the intrinsic states
associated with the bands $g^{\pm}, \beta^{\pm}, \gamma^{\pm}$.
From these states one obtains two sets of angular momentum projected states:
\begin{equation}
\varphi^{(1,\pm)}_{JM}={\cal N}^{(1,\pm)}_JP^J_{M1}\Psi^{(1,\pm)},
\end{equation}
with the projection operator defined by Eq. (\ref{PJMK}).
The dipole projected state can be written in a tensorial form:
\begin{equation}
\varphi^{(1,\pm)}_{JM}={\cal N}^{(1,\pm)}_J
\sum_{J_2,J_3}C^{J_3\;J_2\;J}_{1\;\;\;0\;\;\;1}\left(N^{(\pm)}_{31;J}N^{(g)}_{J_2}\right)^{-1}
\left[\varphi^{(\pm)}_{31,J_3}\varphi^{(g)}_{J_2}\right]_{JM},
\end{equation}
where the octupole factor state is defined by:
\begin{equation}
\varphi^{(1,\pm)}_{31;JM}=N^{(\pm)}_{31;J}P^J_{M1}\Omega^{\dag}_3b^{\dag}_{31}\Psi^{\mp}_{oc}.
\end{equation}
The norm factors  ${\cal N}^{(1,\pm)}_J, N^{(\pm)}_{31;J}$ are analytically
given in Appendix A.

It is worth to mention an useful property of the projected state defined above.
Taking into account the expression of $\vec{J}_3$ in terms of octupole bosons:
\begin{equation}
J^{(3)}_{\mu}=\sqrt{12}\left[b^{\dagger}_3b_3\right]_{1\mu},
\end{equation}
one finds:
\begin{equation}
b^{\dagger}_{31}\Psi^{\mp}_{oc}=\frac{1}{{\cal A}}J^{(3)}_{1}\Psi^{\pm}_{oc},
\end{equation}
where
\begin{equation}
{\cal A}=-\sqrt{12}C^{3\;3\;1}_{1\;0\;1} f.
\end{equation}
Commuting the angular momentum and the rotation operators, one arrives at the following expression for
the octupole projected state:
\begin{equation}
\varphi^{(1,\pm)}_{31;JM}=N^{(\pm)}_{31;J}
\frac{1}{{\cal A}}C^{J^{\prime}\;\; 1\;\;  J}_{0\;\;\; 1\;\;\; 1}
\sum_{\mu, J^{\prime}}C^{J^{\prime}\;\; 1\;\;  J}_{M^{\prime}\;\; \mu\;\; M}
J^{(3)}_{\mu}P^{J^{\prime}}_{M^{\prime}0}\Omega_3\Psi^{\pm}_{oc} .
\label{31JM}
\end{equation}
Denoting the $K^{\pi}=0^{\pm}$ projected state by:
\begin{equation}
\varphi^{(\pm)}_{3;JM}=N^{(\pm)}_{3;J}P^J_{M0}\Omega^{\dag}_3\Psi^{\pm}_{oc} ,
\end{equation}
the equation (\ref{31JM}) leads to:
\begin{equation}
\varphi^{(1,\pm)}_{31;JM}=N^{(\pm)}_{31;J}
\frac{1}{{\cal A}}C^{J\;\; 1\;\;  J}_{0\;\; \;1\;\; \;1}
\sqrt{J(J+1)}\left(N^{(\pm)}_{3;J}\right)^{-1}\varphi^{(\pm)}_{3;JM}.
\end{equation}
Since the two projected states involved in the two sides of the above equation respectively,
are both normalised to unity we have:
\begin{eqnarray}
\varphi^{(1,\pm)}_{31;JM}&=&\varphi^{(\pm)}_{3;JM},\nonumber\\
\left(N^{(\pm)}_{31;J}\right)^{-1}&=&
\frac{1}{{\cal A}}C^{J\;\; 1\;\;  J}_{0\;\;\; 1\;\; \;1}
\sqrt{J(J+1)}\left(N^{(\pm)}_{3;J}\right)^{-1}.
\end{eqnarray}
This equation  provides technical simplifications for calculating the matrix
elements corresponding the dipole projected states.

Invoking the results obtained for the quantum number $K$, one can prove that
the dipole projected states are $K^{\pi}=1^{\pm}$ states, respectively.
For a given J the projected states of positive and negative parity are obviously
orthogonal onto each other. Moreover, they are orthogonal on the states of similar
angular momentum
describing the member states of the six bands which were previously defined.

The dipole projected states are weakly coupled to the states of other bands by
the ${\cal B}_1$ and ${\cal B}_3$ terms of H (2.15). Moreover, these terms give large contribution to the diagonal matrix elements
involving the projected dipole states.
Aiming at describing quantitatively the properties of the dipole states two terms are added to the model Hamiltonian.
\begin{equation}
\Delta H={\cal C}_1\Omega^{\dagger}_3\Omega_3 +
{\cal C}_2\Omega^{\dagger}_3\hat{N}_2\Omega_3.
\end{equation}
The new terms affect only the diagonal matrix elements of the dipole states.Their strengths
are fixed as follows: ${\cal C}_2$ is determined such that the corresponding contribution to the particular state energy, in the negative dipole band,
cancels the one coming from the ${\cal B}_1$ term. ${\cal C}_1$ is fixed such that the measured  excitation energy of the state $1^-$
is reproduced. With the new parameters determined in this way, the effect of
the off diagonal matrix elements corresponding the ${\cal B}_1$  and ${\cal B}_3$ terms,
on the energies in the two dipole bands amounts of a few keV.
Due to this feature the energies of the two dipole bands are obtained as the corresponding average values
of the model Hamiltonian, $H+\Delta H$.

\section{Numerical results}
\renewcommand{\theequation}{4.\arabic{equation}}
\setcounter{equation}{0}

\label{sec:level4}

\subsection{Parameter description}
The formalism presented in the previous section has been numerically applied for seven nuclei:
$^{158}$Gd, $^{172}$Yb, $^{226}$Ra, $^{228}$Th, $^{232}$Th, $^{238}$U, $^{238}$Pu.
Since some results for $^{172}$Yb and $^{226}$Ra were earlier
reported \cite{Rad31,Rad32}, for these nuclei we mention only
the features not presented there. The experimental data are taken from
Refs. \cite{Sug,Blo,Lee} ($^{158}$Gd), \cite{Gono,Gai,Wal,Bal} ($^{172}$Yb),
\cite{Ell,Wol,Coc1}
($^{226}$Ra), \cite{Har} ($^{228}$Th), \cite{Coc1,Sim,Ale,Shu} ( $^{232}$Th ),
\cite{Ale,Shu,Gai} ($^{238}$U), \cite{Shu,Gai,Led} ($^{238}$Pu).
Moreover, three pairs of parity partner bands have been treated in 
Refs.\cite{Rad110,Rad12,Rad13}
where, excepting the new strengths ${\cal C}_1$ and ${\cal C}_2$, all  parameters have been
fixed through the least square procedure.

These new parameters have been fixed as explained in the previous section.
Since the dipole states energies are sensitive to changing
${\cal B}_1$ and ${\cal B}_3$, we change slightly the known values of these
parameters in order to improve the
agreement in the negative dipole band. However, changing the values of
${\cal B}_1$ and ${\cal B}_3$ affects some of calculated energies in the 
other six bands.
 Such corrections
are washed out by a small change of one of the parameters ${\cal A}_J, {\cal A}_{J23}$.
 We have checked for few cases that the results obtained in this way are
 similar to those provided by
a least square procedure applied for all eight bands. The final results for the
model parameters are listed in Table I.

\begin{table}
\begin{tabular}{|c|c|c|c|c|c|c|c|}
\hline
  &  $^{158}$Gd & $^{172}$Yb  & $^{226}$Ra & $^{228}$Th & $^{232}$Th &  $^{238}$U &  $^{238}$Pu   \\
\hline
d &  3.0        &   3.68      &  3.0       &  3.1       &  3.25      &    3.9     &   3.9          \\
f &  0.3        &   0.3       &  0.8       &  0.3       &  0.3       &    0.6     &   0.3          \\
${\cal A}_1$&21.49& 26.94      & 20.29      &  17.72     &  14.26     &    20.64   & 18.84\\
${\cal A}_2$&-12.28&-17.68     &-17.21      & -12.67     &  -8.34     &   -9.59    & -8.63\\
${\cal A}_J$&3.5  &  4.72      &  0.49      &  1.32      &   2.26     &    2.14    &  2.26\\
${\cal A}_{J23}$&15.0&4.70     &7.17        &  8.38      &   6.00     &    5.00    &  5.00\\
${\cal B}_1$&-11.68  &-24.29   &-1.53       & -2.79      &  -6.25     &  -11.97    & -8.43\\
${\cal B}_3$&3414.62 & 8327.68 &523.07      & 858.37     &2047.04     & 4483.28    & 3254.09\\
${\cal C}_1$& -3096.52&-8853.24 & -217.31   & -603.58    &-1879.04    &-4663.88    &-3265.85\\
${\cal C}_2$&  285.93 &594.45   & 38.18     & 68.16      & 152.87     &  295.64    & 206.29  \\
\hline
\end{tabular}
\caption{The deformation parameters $d$ and $f$ and the structure coefficients
involved in the model
Hamiltonian, obtained as described in the text, are listed for several
isotopes.
The deformations are dimensionless, while the remaining coefficients are given
in units of keV.}
\label{Table I}
\end{table}

\subsection{Dipole bands energies}

As we have already suggested before, the calculated energies for
$g^{\pm}, \beta^{\pm}, \gamma^{\pm}$ are practically the same as in 
Ref.\cite{Rad110,Rad12,Rad13}
 and therefore they are not given here.
We stress on the fact that the volume of explained data with the mentioned
parameters is quite large.
For example, in the  previously treated six bands of $^{232}$Th,
about 55 excitation energies are known.
Also, with the fixed deformation parameters,  several experimental
data concerning
the transition reduced probabilities are realistically  described. It is interesting to mention that these
parameters have specific dependence on A and Z
which means that applying the formalism to new cases, the strength parameters
can be considered as fully determined from the previous analysis.
As shown  in Figs. 1 and 2, the new parameters ${\cal C}_1$ and ${\cal C}_2$ exhibit also a smooth dependence on
the variable $A-0.5(N-Z)$. Adding the third isospin component to A we avoided
the situation when for the isotopes of the same A one obtains different values
of the considered parameters, which results in having a ill-defined function.
The calculated energies for the dipole bands are collected in Tables II-V.  Only the states with angular momentum not larger than
20 are listed. Note that except for $^{172}$Yb, only few data are known for these bands.
From the energy analysis, several common features can be seen.  We note that 
in both $K^{\pi}=1^-$ and
$K^{\pi}=1^+$ bands a doublet structure shows up. For us it is not clear whether this doublet structure is an indication
of two bands of odd and even spins respectively. This suspicion is somehow confirmed in
$^{228}$Th and $^{226}$Ra, where in the low part of the spectrum the doublet members have
not a natural energy ordering.

\begin{table}
\begin{tabular}{|c|cc|cc|cc|cc|}
\hline
& \multicolumn{2}{c}{$^{158}$Gd}& \multicolumn{2}{c}{$^{172}$Yb}& \multicolumn{2}{c}{$^{228}$Th}& \multicolumn{2}{c|}{$^{232}$Th}\\
\hline
J$^+$ &  Exp. & Th. &  Exp. & Th. &   Exp. & Th. &   Exp.& Th. \\
\hline
1$^+$ &2.534  &2.531&2.010  &1.880&        &1.247&1.489  &1.508\\
2$^+$ &2.539  &2.563&2.047  &1.897&        &1.250&       & 1.519\\
3$^+$ &2.631  &2.638&       &1.970&        &1.311& 1.561 & 1.567\\
4$^+$ &       &2.698&2.073  &1.984&        &1.301& 1.573 & 1.578\\
5$^+$ &       &2.829&       &2.133&        &1.426&       & 1.673\\
6$^+$ &       &2.927&2.156  &2.139&        &1.409&       & 1.687\\
7$^+$ &       &3.108&       &2.368&        &1.595&       & 1.827\\
8$^+$ &       &3.256&       &2.370&        &1.587&       & 1.851\\
9$^+$ &       &3.470&       &2.676&        &1.818&       & 2.029\\
10$^+$&       &3.681&       &2.683&        &1.832&       & 2.073\\
11$^+$&       &3.911&       &3.056&        &2.092&       & 2.276\\
12$^+$&       &4.189&       &3.078&        &2.139&       &  2.347\\
13$^+$&       &4.423&       &3.506&        &2.412&       & 2.567\\
14$^+$&       &4.771&       &3.552&        &2.497&       & 2.669\\
15$^+$&       &5.001&       &4.023&        &2.774&       & 2.899\\
16$^+$&       &5.417&       &4.101&        &2.899&       & 3.033\\
17$^+$&       &5.636&       &4.606&        &3.174&       & 3.628\\
18$^+$&       &6.118&       &4.719&        &3.338&       & 3.435\\
19$^+$&       &6.323&       &5.251&        &3.606&       & 3.672 \\
20$^+$&       &6.870&       &5.403&        &3.808&       & 3.871\\
\hline
\end{tabular}
\caption{Experimental (left column) and theoretical (right column) excitation
energies for the dipole band $K^{\pi}=1^+$ states are given in units of MeV for
the isotopes $^{158}Gd, ^{172}Yb, ^{228}Th, ^{232}Th$.}
\label{Table II}

\end{table}

\begin{table}
\begin{tabular}{|c|cc|cc|cc|}
\hline
& \multicolumn{2}{c}{$^{226}$Ra}& \multicolumn{2}{c}{$^{238}$U}& \multicolumn{2}{c|}{$^{238}$Pu}\\
\hline
J$^+$ &  Exp. & Th. &  Exp. & Th. &   Exp. & Th.  \\
\hline
1$^+$ &       &1.363&1.354  &1.367&  1.310 &1.343\\
2$^+$ &       &1.345&       &1.380&        &1.357 \\
3$^+$ & 1.422 &1.420&       &1.425&        & 1.401\\
4$^+$ &       &1.359&       &1.442&        & 1.420\\
5$^+$ &       &1.526&       &1.531&        & 1.506\\
6$^+$ &       &1.432&       &1.546&        &1.525 \\
7$^+$ &       &1.684&       &1.684&        &1.656 \\
8$^+$ & 1.587 &1.582&       &1.698&        &1.677\\
9$^+$ &       &1.896&       &1.882&        &1.851 \\
10$^+$&       &1.806&       &1.901&        &1.879 \\
11$^+$&       &2.158&       &2.126&        &2.092 \\
12$^+$&       &2.094&       &2.156&        &2.132\\
13$^+$&       &2.465&       &2.413&        &2.376 \\
14$^+$&       &2.433&       &2.460&        &2.433\\
15$^+$&       &2.812&       &2.743&        &2.702 \\
16$^+$&       &2.812&       &2.811&        &2.780 \\
17$^+$&       &3.193&       &3.113&        &3.067 \\
18$^+$&       &3.223&       &3.205&        &3.171\\
19$^+$&       &3.602&       &3.521&        &3.471\\
20$^+$&       &3.662&       &3.639&        &3.601\\
\hline
\end{tabular}
\caption{The same as in Table II but for $^{226}Ra, ^{238}U, ^{238}Pu$.}
\label{Table III}

\label{Table V}
\end{table}
\begin{table}
\begin{tabular}{|c|cc|cc|cc|cc|}
\hline
& \multicolumn{2}{c}{$^{158}$Gd}& \multicolumn{2}{c}{$^{172}$Yb}& \multicolumn{2}{c}{$^{228}$Th}& \multicolumn{2}{c|}{$^{232}$Th}\\
\hline
J$^-$ &  Exp. & Th. &  Exp. & Th. &   Exp. & Th. &   Exp.& Th. \\
\hline
1$^-$ &1.856  &1.856&1.155  &1.155&  0.952 &0.952&1.078  &1.078\\
2$^-$ &1.895  &1.912&1.198  &1.207&  0.968 &0.989&1.100  &1.110 \\
3$^-$ &1.978  &1.970&1.222  &1.257&        &1.017&       &1.140 \\
4$^-$ &       &2.091&1.331  &1.375&        &1.104&       &1.213 \\
5$^-$ &       &2.184&1.353  &1.443&        &1.142&       &1.257 \\
6$^-$ &       &2.366&1.541  &1.636&        &1.281&       &1.372 \\
7$^-$ &       &2.500&1.567  &1.716&        &1.330&       &1.428 \\
8$^-$ &       &2.727&1.828  &1.986&        &1.512&       &1.582 \\
9$^-$ &       &2.911&1.849  &2.077&        &1.578&       &1.654 \\
10$^-$&       &3.165&2.193  &2.421&        &1.791&       &1.839 \\
11$^-$&       &3.404&2.209  &2.524&        &1.880&       &1.931 \\
12$^-$&       &3.672&2.630  &2.935&        &2.113&       &2.140  \\
13$^-$&       &4.970&2.646  &3.053&        &2.227&       &2.255 \\
14$^-$&       &4.237&3.134  &3.523&        &2.470&       &2.479 \\
15$^-$&       &4.598&       &3.661&        &2.613&       &2.619 \\
16$^-$&       &4.857&       &4.182&        &2.860&       &2.854 \\
17$^-$&       &5.281&       &4.342&        &3.033&       &3.021 \\
18$^-$&       &5.526&       &4.906&        &3.277&       &3.261 \\
19$^-$&       &6.013&       &5.091&        &3.481&       &3.456  \\
20$^-$&       &6.240&       &5.692&        &3.719&       &3.699 \\
\hline
\end{tabular}
\caption{ The same as in Table II, but for the $K^{\pi}=1^-$ band.}
\label{Table IV}

\end{table}

\begin{table}
\begin{tabular}{|c|cc|cc|cc|}
\hline
& \multicolumn{2}{c}{$^{226}$Ra}& \multicolumn{2}{c}{$^{238}$U}& \multicolumn{2}{c|}{$^{238}$Pu}\\
\hline
J$^-$ &  Exp. & Th. &  Exp. & Th. &   Exp. & Th.  \\
\hline
1$^-$ &1.080  &1.049&0.967  &0.967&  0.963 &0.863\\
2$^-$ &1.102  &1.090&0.988  &0.998&  0.986 &0.992 \\
3$^-$ &       &1.108&1.035  &1.033&  1.019 &1.025 \\
4$^-$ &       &1.211&1.053  &1.100&  1.083 &1.089 \\
5$^-$ &       &1.227&       &1.153&        &1.138 \\
6$^-$ &       &1.394&       &1.259&        &1.240 \\
7$^-$ &       &1.409&       &1.326&        &1.302 \\
8$^-$ &       &1.631&       &1.472&        &1.443\\
9$^-$ &       &1.647&       &1.553&        &1.516 \\
10$^-$&       &1.913&       &1.736&        &1.695 \\
11$^-$&       &1.933&       &1.831&        &1.781 \\
12$^-$&       &2.233&       &2.048&        &1.994  \\
13$^-$&       &2.258&       &2.157&        &2.093 \\
14$^-$&       &2.584&       &2.405&        &2.336\\
15$^-$&       &2.619&       &2.529&        &2.450 \\
16$^-$&       &2.962&       &2.802&        &2.719 \\
17$^-$&       &3.012&       &2.942&        &2.851 \\
18$^-$&       &3.365&       &3.237&        & 3.140\\
19$^-$&       &3.434&       &3.392&        &3.291  \\
20$^-$&       &3.790&       &3.707&        &3.597 \\
\hline
\end{tabular}
\caption{ The same as in Table III but for the $K^{\pi}=1^-$ band.}
\end{table}

The excitation energies were further used to represent, in Figs. 3-7, the dynamic moment of inertia as function
of angular frequency defined as:

\begin{eqnarray}
\hbar\omega_I&=&\frac{dE}{dI}\approx \frac{1}{2}(E_I-E_{I-2}),\nonumber\\
\frac{{\cal J}^{(2)}}{\hbar^2}&=&\left(\frac{d\hbar\omega}{dI}\right)^{-1}
\approx \frac{2}{\hbar (\omega_I-\omega_{I-2})}.
\end{eqnarray}
The common feature of the moments of inertia is the zigzag structure in both, the negative and positive parity bands.
For the $1^+$ band, the moments of inertia of odd and even spins are lying on smooth curves, respectively. The curve for the odd spins
lies above that of even spins. The same is true also for the negative dipole
band with the difference that the
curve corresponding to the even angular momenta is higher than that for odd values of angular momentum.
Due to the relative position of the four curves comprising the moments of inertia
of even and odd spin states of positive and negative parity respectively,
for some nuclei ($^{172}$Yb, $^{226}$Ra, $^{238}$U and $^{238}$Pu)
it turns out that for some ranges of angular momenta  the (odd,positive);(even, negative) and (even,positive); (odd,negative)
states have moments of inertia lying on similar curves, respectively.
This interleaved structure might be a signature for an octupole deformation in these states.
In order get a confirmation for such an expectation we plotted in the low panels of the above quoted figures the first and second order
energy displacement functions defined as:
\begin{eqnarray}
\delta E(J^-)&=&E(J^-)-\frac{(J+1)E((J-1)^+)+JE((J+1)^+)}{2J+1},
\nonumber\\
\Delta E_{1,\gamma}(J)&=&\frac{1}{16}\left[6E_{1,\gamma}(J)-4E_{1,\gamma}(J) -
4E_{1,\gamma}(J) +E_{1,\gamma}(J) +E_{1,\gamma}(J) \right],\nonumber\\
E_{1,\gamma}(J)&=&E(J+1)-E(J).
\end{eqnarray}
If the parity partner bands have similar J(J+1) pattern in a certain range of
angular momentum, then
the function $\delta E$ is vanishing for I belonging to the mentioned range. The reverse
statement, if valid, asserts that for the angular momenta where the first
order displacement function
vanishes, the partner bands have identical moments of inertia which further 
infer that
the two bands can be viewed as being associated to a sole intrinsic state.
However, the J dependence of the excitation energies for the considered nuclei
deviates
from the J(J+1) law. If the energies can be described by a second order
polynomial in $J(J+1)$
and moreover, the partner bands are characterised by the same strength for
the $\left[J(J+1)\right]^2$ term, the second order energy displacement function is vanishing.
Reversely, if  $\Delta E_{1,\gamma}$ is vanishing, this is a sign that the two
partner bands have similar
$\left[J(J+1)\right]^2$   pattern.
Concerning the second order energy displacement function, one should mention that
there are two distinct functions of angular momentum differing by the set of states involved.
In one function the lowest state is $1^+$ (the black squares) while for the other function
the state $1^-$ is the lowest in energy. The parity assignment for the states involved in $\Delta E$ is
conventionally taken as follows. The states whose angular momenta differ by two units have the same parity
while those which differ by unity are of different parities.
Inspecting Figs. 6, 7 from the present paper and
3 of Refs.(\cite{Rad31,Rad32}) we remark that for $^{172}$Yb, $^{238}$U, and $^{238}$Pu the second order
energy displacements vanish for 2-3 consecutive values of angular momenta, while for
$^{226}$Ra this is zero or very close to zero for $I\geq 11$.

In the right upper corner of Figs. 3-7, we plotted the angle between the angular momenta carried by the quadrupole and octupole bosons
, respectively in the dipole states of positive as well as of negative parity.
Such angle is defined as:
\begin{equation}
\cos\varphi=\frac{\langle\varphi^{(k)}_{JM}|\vec{J}_2\cdot \vec{J}_3|\varphi^{(k)}_{JM}\rangle}
{\sqrt{\langle\varphi^{(k)}_{JM}|\hat{J}^2_2|\varphi^{(k)}_{JM}\rangle
\langle\varphi^{(k)}_{JM}|\hat{J}^2_3|\varphi^{(k)}_{JM}\rangle}},\;\;k=1,+;1,-.
\end{equation}
Note that this angle is a decreasing function of angular momentum and that
the angles for odd and even spin states of positive parity, respectively are lying on
smooth curves. The same is true for the angles characterising the negative parity band.
Moreover, for $I\leq 7$ the curves for odd spin states of positive parity  and for
even spin states of negative parity are very close to each other.
The same is valid for the curves of even spin and positive parity and odd spin states of negative parity.
Similarly, one could calculate the angle between the two angular momenta in
the other parity partner bands. Here we give the results for the bands $0^+$
and $0^-$ in Figs. 8 and 9. For these bands we didn't consider the admixture
with the gamma band states of similar angular momenta,
since the numerical results for the isotopes considered, the mixing amplitudes
are small.
For a better presentation we omitted the state $0^+$ where the angle is equal
to $\pi$.
The angles for the two bands exhibit minima which are achieved for different values of angular momenta.
        However, for $^{226}$Ra and $^{238}$U the two minima are almost equal
        to each other and
        are reached for close values of angular momenta. After reaching the minima the angles
        increase and approach the limit value of $\frac{\pi}{2}$ in both bands.
In the remaining cases this limit is met first by the band $0^-$ and much later
in the ground band.  Let us comment on the states where the angular momenta
determined by the quadrupole and octupole bosons respectively, are perpendicular on
each other respectively. The system under such a state constitute a precursor
of a chiral symmetry \cite{Frau1}.
Indeed, we could imagine a system of nucleons moving around a phenomenological
core described by the quadrupole-octupole boson Hamiltonian considered here.
Suppose that the coupling of the particle and core subsystems is such that the
angular momentum carried by particles, say $\vec{j}$, is perpendicular to the plane
$(\vec{J}_2,\vec{J}_3)$. If the system energy corresponding to the situation
when the set $(\vec{j},\vec{J}_2,\vec{J}_3)$ form a right triad is degenerate with the energy corresponding
to the situation when the three vectors define a left triad, one says that the system has a chiral symmetry.
Of course, such a situation is an ideal picture and in practice one expects that the two energies are only
approximatively degenerate. The symmetry breaking is expected to yield some
 properties which are specific for the new nuclear phase.
\subsection{Electromagnetic transition probabilities}
The E1 and M1 transitions are determined by the following transition operators:

\begin{eqnarray}
E_{1\mu}&=&T^{(h)}_{1\mu}+T^{(anh)}_{1\mu},\nonumber\\
T^{(h)}_{1\mu}&=&q_h\sum_{\mu_2,\mu_3}C^{3\;\;2\;\;1}_{\mu_3\;\mu_2\;\mu}
\left(b^{\dagger}_{3\mu_3}+(-)^{\mu_3}b_{3,-\mu_3}\right)
\left(b^{\dagger}_{2\mu_2}+(-)^{\mu_2}b_{2,-\mu_2}\right),\nonumber\\
T^{anh}_{1\mu}& =& \left[b^{\dag}_3\left(\hat{J}_3\hat{J}_2\right)\right]_{1\mu}+
\left[\left(\hat{J}_2\hat{J}_3\right)b_3\right]_{1\mu},
\nonumber\\
M_{1\mu}&=&g_2\left(\hat{J}_{2}\right)_{\mu}+g_3\left(\hat{J}_{3}\right)_{\mu}
\nonumber\\
&+&
g^{\prime}_2\left[\left(\hat{J}_{2}\left(b^{\dag}_3b^{\dag}_3\right)_2\right)_{1\mu}+
\left(\left(b_3b_3\right)_2\hat{J}_2\right)_{1\mu}\right]
\nonumber\\
&+&
g^{\prime}_3\left[\left(\hat{J}_{3}\left(b^{\dag}_2b^{\dag}_2\right)_2\right)_{1\mu}+
\left(\left(b_2b_2\right)_2\hat{J}_3\right)_{1\mu}\right] .
\label{E1andM1}
\end{eqnarray}

The reduced matrix elements $^{*)}$\footnote{$^{*)}$Throughout this paper the Rose convention
for the Wigner Eckardt theorem is used \cite{Rose}} of interest for these
operators are given analytically
 in Appendix C.
 Let us first discuss the magnetic properties of the dipole bands.
 Firstly, we calculated the gyromagnetic factors for the states of the two bands by
considering only the lowest order boson terms in the expression of the $M1$ transition operator.
In Ref.\cite{Rad200} we derived an expression for the M1 transition operator by quantising its classical expression.
The important result was that the gyromagnetic factors $g_2$ and $g_3$
were expressed in terms of the Hamiltonian structure coefficients.The values obtained for
$^{238}$U are:
\begin{equation}
g_2=0.371 \mu_N,\;\; g_3=2.266 \mu_N.
\end{equation}
These values have been adopted for all nuclei considered here.
The results are presented in Tables VI and VII. We remark  that the gyromagnetic factor of the state
$2^-$ is very close to the phenomenologically adopted value for nuclei in the
ground state, i.e. $Z/A$. This value is met in the positive parity band for
the state $13^+$. The gyromagnetic factor of even spin states of positive parity
is constantly much larger than those of negative parity. By contrary, the odd spin states of positive and negative parity
have close gyromagnetic factors. For $J\leq 5$ the odd spin states of positive parity have gyromagnetic factors which are slightly larger than those characterising the odd spin states of negative parity.
Starting with $J=7$, the ordering of gyromagnetic factors of odd spin states in the two bands is changed.

\begin{table}
\begin{tabular}{|c|cc|cc|cc|cc|}
\hline
& \multicolumn{2}{c}{$^{158}$Gd}& \multicolumn{2}{c}{$^{172}$Yb}& \multicolumn{2}{c}{$^{228}$Th}& \multicolumn{2}{c|}{$^{232}$Th}\\
\hline
J & $K^{\pi}=1^+$ & $K^{\pi}=1^-$ &  $K^{\pi}=1^+$ & $K^{\pi}=1^-$ &   $K^{\pi}=1^+$ & $K^{\pi}=1^-$ &   $K^{\pi}=1^+$& $K^{\pi}=1^-$ \\
\hline
1 &0.645  &0.865  &0.645  &0.789   &0.645   &0.851    &0.645  &0.832\\
2 &1.081  &0.403  &0.859  &0.379   &1.042   &0.399    &0.989  &0.392 \\
3 &0.503  &0.548  &0.424  &0.452   &0.489   &0.530    &0.469  &0.506 \\
4 &0.939  &0.316  &0.760  &0.286   &0.910   &0.310    &0.869  &0.302 \\
5 &0.476  &0.486  &0.395  &0.405   &0.461   &0.472    &0.441  &0.452 \\
6 &0.821  &0.293  &0.697  &0.263   &0.803   &0.386    &0.776  &0.280 \\
7 &0.457  &0.448  &0.381  &0.383   &0.444   &0.437    &0.425  &0.422 \\
8 &0.723  &0.281  &0.642  &0.254   &0.712   &0.276    &0.695  &0.269 \\
9 &0.439  &0.417  &0.372  &0.366   &0.428   &0.409    &0.412  &0.398 \\
10&0.645  &0.273  &0.592  &0.249   &0.638   &0.269    &0.627  &0.263 \\
11&0.422  &0.391  &0.363  &0.352   &0.412   &0.386    &0.398  &0.377 \\
12&0.584  &0.267  &0.548  &0.245   &0.579   &0.263    &0.573  &0.258  \\
13&0.406  &0.370  &0.355  &0.340   &0.398   &0.366    &0.386  &0.359 \\
14&0.536  &0.261  &0.511  &0.241   &0.533   &0.258    &0.528  &0.253 \\
15&0.391  &0.353  &0.347  &0.329   &0.384   &0.349    &0.374  &0.344 \\
16&0.498  &0.256  &0.480  &0.238   &0.496   &0.253    &0.492  &0.249 \\17&0.376  &0.338  &0.339  &0.319   &0.371   &0.336    &0.362  &0.331 \\
18&0.476  &0.252  &0.453  &0.237   &0.466   &0.250    &0.463  &0.246 \\
19&0.364  &0.327  &0.332  &0.310   &0.359   &0.324    &0.352  &0.321 \\
20&0.442  &0.249  &0.431  &0.234   &0.440   &0.246    &0.438  &0.243 \\
\hline
\end{tabular}
\caption{Calculated gyromagnetic factors for the states belonging to the
dipole $K^{\pi}=1^{+}$  (left column)
and
$K^{\pi}=1^{-}$ (right column) bands are given in units of nuclear magneton
$\mu_N$, for $^{158}Gd, ^{172}Yb, ^{228}Th, ^{232}Th$.}
\label{Table VI}
\end{table}

\begin{table}
\begin{tabular}{|c|cc|cc|cc|}
\hline
& \multicolumn{2}{c}{$^{226}$Ra}& \multicolumn{2}{c}{$^{238}$U}& \multicolumn{2}{c|}{$^{238}$Pu}\\
\hline
J & $K^{\pi}=1^+$ & $K^{\pi}=1^-$ &  $K^{\pi}=1^+$ & $K^{\pi}=1^-$ &   $K^{\pi}=1^+$ & $K^{\pi}=1^-$ \\
\hline
1 &0.645  &0.931  &0.645  &0.793   & 0.645  &0.773  \\
2 &1.100  &0.427  &0.813  &0.381   &0.806   &0.374   \\
3 &0.510  &0.644  &0.409  &0.461   &0.407   &0.431   \\
4 &0.959  &0.351  &0.719  &0.289   &0.712   &0.279   \\
5 &0.484  &0.592  &0.378  &0.417   &0.375   &0.385   \\
6 &0.841  &0.334  &0.667  &0.268   &0.659   &0.257   \\
7 &0.467  &0.563  &0.366  &0.398   &0.363   &0.365   \\
8 &0.745  &0.329  &0.621  &0.260   &0.614   &0.248  \\
9 &0.451  &0.544  &0.358  &0.386   &0.355   &0.352   \\
10&0.670  &0.330  &0.579  &0.255   &0.572   &0.243   \\
11&0.437  &0.531  &0.351  &0.377   &0.347   &0.340   \\
12&0.614  &0.332  &0.542  &0.253   &0.534   &0.239    \\
13&0.423  &0.522  &0.344  &0.369   &0.341   &0.330   \\
14&0.572  &0.337  &0.509  &0.252   &0.500   &0.236   \\
15&0.412  &0.514  &0.338  &0.362   &0.334   &0.320   \\
16&0.541  &0.342  &0.480  &0.251   &0.472   &0.234  \\
17&0.403  &0.507  &0.332  &0.356   &0.328   &0.312   \\
18&0.518  &0.348  &0.457  &0.250   &0.447   &0.232   \\
19&0.395  &0.499  &0.327  &0.351   &0.322   &0.304   \\
20&0.500  &0.354  &0.436  &0.250   &0.426   &0.230   \\
\hline
\end{tabular}
\caption{Gyromagnetic factors for the states belonging to the
dipole $K^{\pi}=1^{-}$ and
$K^{\pi}=1^+$ bands.}
\label{Table VII}
\end{table}
The transition from the band $1^+$ to the ground band is caused by the 
anharmonic term of the
transition operator, while the intraband transitions as well as the 
gyromagnetic factors have been calculated
by using only the lowest order boson terms. The factors $g_2$ and $g_3$ have
been
taken as mentioned before.
Therefore the branching ratios for the M1 transitions:
\begin{eqnarray}
R^{10}_{++}&=&\left[\frac{\langle \varphi^{(1,+)}_{J}||M_1||\varphi^{(g,+)}_{J+1}\rangle}
{\langle \varphi^{(1,+)}_{J}||M_1||\varphi^{(g,+)}_{J-1}\rangle}\right]^2,\nonumber\\
R^{11}_{++}&=&\left[\frac{\langle \varphi^{(1,+)}_{J}||M_1||\varphi^{(1,+)}_{J+1}\rangle}
{\langle \varphi^{(1,+)}_{J}||M_1||\varphi^{(1,+)}_{J-1}\rangle}\right]^2,\nonumber\\
R^{11}_{--}&=&\left[\frac{\langle \varphi^{(1,-)}_{J}||M_1||\varphi^{(1,-)}_{J+1}\rangle}
{\langle \varphi^{(1,-)}_{J}||M_1||\varphi^{(1,-)}_{J-1}\rangle}\right]^2,
\end{eqnarray}
are free of any adjustable parameter. The calculated values for these ratios are
 given in Tables VIII and IX. The branching ratios to the ground band 
have a
 minimum for $7\leq J\leq 9$ and a maximum for $15\leq J\leq 15$.
Exceptions are for $^{238}$U and $^{238}$Pu where the maximum values are reached for J=23.
The dominant ratios are those for odd values of angular momentum.
The same is true for the intraband transition for the band $1^+$.

By contrast, in the negative parity band the ratios corresponding to even
angular momenta prevail.
One notices that for $^{158}$Gd, $R^{11}_{++}$ has a minimum value for $J=9$ while
$R^{11}_{--}$ has a maximum for $J=8$. These extreme values change from one nucleus to another.
The dominant intraband M1 transitions for the band $1^+$ are those from even
spin states.
Moreover, they increase with the angular momentum. For example, for $^{232}$Th
the B(M1) value
is 0.25$\mu_n^2$ for $J=2$  and 4.23$\mu_N^2$ for $J=30$.
As for the band $1^-$ the dominant transitions are those from odd spin states.
Indeed, for the isotope mentioned above the $B(M1)$ value increase from 0.45 for J=3
to 2.06 $\mu_N^2$ for $J=29$. Except for the first transitions ($2^+\to 1^+$)
all others B(M1) values are larger than the ones associated to  negative parity band.
Due to these facts we say that the band $1^+$ has a dominant magnetic character.
It is worth noting that while the collective magnetic states of scissors nature are determined by the angular vibration, in a scissors fashion, of the symmetry axis of the proton and neutron systems, that angle being quite small, here the angle between $\vec{J}_2$ and $\vec{J}_3$ (which might be assimilated with the angle between the axis of the maximal moments of inertia of the quadrupole and octupole systems, respectively) is large. In this respect we could call the magnetic states from the band $1^+$, shares like states. 
\begin{table}
\begin{tabular}{|c|ccc|ccc|ccc|ccc|}
\hline
& \multicolumn{3}{c}{$^{158}$Gd}& \multicolumn{3}{c}{$^{172}$Yb}& \multicolumn{3}{c}{$^{228}$Th}& \multicolumn{3}{c|}{$^{232}$Th}\\
\hline
J & $1^+\to 0^+$ & $1^+\to 1^+$ &$1^-\to 1^-$ &$1^+\to 0^+$ & $1^+\to 1^+$ &$1^-\to 1^-$ &$1^+\to 0^+$ & $1^+\to 1^+$ &$1^-\to 1^-$ & $1^+\to 0^+$ & $1^+\to 1^+$ &$1^-\to 1^-$\\
\hline
1 &0.376  &       &       &0.369   &        &       &0.375  &        &         & 0.373   &   & \\
2 &       &       &7.718  &        &        &4.272  &       &        &6.819    &         &   &5.838  \\
3 &0.348  &11.51  &       &0.343   &639.1   &       &0.347  &15.74   &         &0.346   & 27.233 &     \\
4 &       &       &20.45  &        &        &6.335  &       &        &15.802   &   &     &11.561         \\
5 &0.160  &5.22  &        &0.185   &21.14   &       &0.164  & 6.144 &          & 0.170   &8.015    &       \\
6 &       &      &97.87  &         &        &12.495 &       &       &58.746    &   &     &33.428         \\
7 &0.004&4.17    &       &0.037   & 9.755  &        &0.008  & 4.613 &          &0.015  &5.446    &         \\
8 &     &        &2326.  &        &        & 27.790 &       &       &465.4     &   &     &139.73           \\
9 &0.300  &4.00  &       &0.024   &6.998   &        &0.214  &4.262  &          &0.128    &4.743    &          \\
10&       &      & 978.4 &        &        &72.40   &       &       &8901.     &   &     &2025.            \\
11&3.114  &4.15 &        & 0.478  &5.979   &        &2.275  & 4.302 &          &1.466    & 4.595   &             \\
12&       &       &223.83  &      &        &258.3   &       &       & 410.81   &         &         & 2105.            \\
13&23.821  &4.45  &      & 2.535  &5.574  &         &15.494  &4.524  &         &8.870   & 4.691   &             \\
14&        &      &122.4  &       &       &2548     &       &       &172.680   &         &         & 353.6        \\
15&549.2  &4.84  &       &11.065   &5.455   &       &189.61  & 4.848 &         & 66.498  &4.916    &                \\
16&       &      &87.86  &        &        & 6013.   &      &        &110.845  &         &         &  173.6                    \\
17&506.6  &5.28  &       &60.088   &5.493   &       & 1629. & 5.238 &          &1018.  &5.220    &               \\
18&       &      &71.37  &         &        &680.7  &       &       &84.690    &         &         & 116.47               \\
19&108.7  & 5.77 &       &1442. &5.627      &        &146.60  &5.670  &        &281.311   &5.576    &                \\
20&       &      &61.92  &       &         &294.14   &      &       &70.747   &         &       & 90.154         \\
21&60.94  &6.28  &       &624.06   &5.825   &         &70.577&6.133  &         &94.363   &5.968    &         \\
22&       &      &55.79  &       &         &181.68    &      &       &62.177   &         &         & 75.443        \\
23&40.07  &6.81  &       &138.12   &6.066   &         &48.986  & 6.617 &       &57.697   &6.386    &         \\
24&       &      &51.42  &        &         &132.16    &     &        &56.352  &         &         & 66.146        \\
25&37.76  &7.36  &       &73.40   &6.345   &          &39.639  &7.118  &       &43.756   &6.825    &         \\
26&       &      &48.06  &        &        &105.31    &      &       &52.066   &         &         & 59.725        \\
27&33.71  &7.90  &       &49.53   &6.598   &          &34.548  & 7.618 &       &36.520   &7.259    &          \\
28&       &      &45.68  &        &        &91.05    &        &       & 49.189  &        &         &55.694           \\
29&31.07  &8.41  &       &37.29   &6.768   &         &31.257  & 8.082 &         &31.951   & 7.649   &           \\
30&       &      & 5.171 &        &        &6.167    &        &       & 5.338  &    &    &   5.568        \\
\hline
\end{tabular}
\caption{M1 branching ratios for the $K^{\pi}=1^{+}$ and $K^{\pi}=1^{-}$ bands
 for $^{158}$Gd, $^{172}$Yb, $^{228}$Th, $^{232}$Th.}
\label{Table VIII}
\end{table}

\begin{table}
\begin{tabular}{|c|ccc|ccc|ccc|}
\hline
& \multicolumn{3}{c}{$^{226}$Ra}& \multicolumn{3}{c}{$^{238}$U}& \multicolumn{3}{c|}{$^{238}$Pu}\\
\hline
J & $1^+\to 0^+$ & $1^+\to 1^+$ &$1^-\to 1^-$ &$1^+\to 0^+$ & $1^+\to 1^+$ &$1^-\to 1^-$ &$1^+\to 0^+$ & $1^+\to 1^+$ &$1^-\to 1^-$ \\
\hline
1 &0.376  &         &       & 0.368  &        &       & 0.368 &        &         \\
2 &       &         &15.218 &        &        &4.408  &       &        &  3.821        \\
3 &0.350  &10.097   &       & 0.344  &2820.   &       &0.343  &1322.       &            \\
4 &       &         & 222.04 &       &        &6.857  &       &        & 5.144   \\
5 &0.165  &4.760    &        & 0.192 &37.780  &       &0.190  &42.575       &       \\
6 &       &         &152.144 &       &        &14.929 &       &       & 9.061           \\
7 &0.007  &3.819   &         & 0.050 &13.244  &       &0.048  &14.215       &     \\
8 &       &        &29.201   &       &        &41.279 &       &       &17.435       \\
9 &0.220  &3.640   &         & 0.006 &8.489   &       &0.007  &8.989       &           \\
10&       &        &14.793   &       &        &190.47 &       &       & 36.714       \\
11&2.315  &3.717   &         & 0.263 &6.769   &       &0.283  &7.135        &          \\
12&       &        &10.067   &       &        &1713. &       &   &88.845      \\
13&15.399  &3.901  &         & 1.412 &6.005   &       & 1.508 &6.323       &          \\
14&        &       & 7.908   &       &        &416.9  &       &       &282.24\\
15&171.476  &4.127 &         & 5.457 &5.652   &       &5.888  &5.959       &     \\
16&        &       &6.768    &       &        &113.8  &       &      &1951  \\
17&2646.  & 4.365  &         & 21.402&5.511   &       &23.781 &5.827      &            \\
18&       &        &6.134    &       &        &58.266 &       &       &   7584       \\
19&171.74  & 4.596 &         & 123.16&5.491   &       &149.878&5.832      &                \\
20&        &       &5.786    &       &        &37.904 &       &       &  1090.      \\
21&79.553  &4.815  &         &1569.   &5.547   &       &1153.  &5.923      &         \\
22&        &       &5.612     &      &        &27.941 &       &       & 411.09   \\
23&54.482  & 5.017  &        & 443.44&5.651   &       &355.806& 6.073      & \\
24&        &       &  5.552  &       &        &22.205 &       &        & 237.69    \\
25&43.842  &5.207   &        & 131.47&5.797   &       &118.581&6.277      &    \\
26&        &       &5.566     &      &        &18.540 &       &       & 165.70  \\
27&38.210  &5.383   &         & 68.75&5.896   &       &64.481 &6.430       &     \\
28&        &        &5.636    &      &        &16.357 &       &       & 132.42     \\
29&34.744  & 5.543  &         &44.709&5.909   &       &42.722 &6.475        &        \\
30&        &        &0.469     &     &        &1.318  &       &       &  6.480   \\
\hline
\end{tabular}
\caption{The same as in Table VIII but for $^{226}$Ra, $^{238}$U, $^{238}$Pu.}
\label{Table IX}
\end{table}
Comparing the values of $R^{11}_{++}$ with those describing the M1 branching
ratios for the transitions relating the bands $1^-$ and $0^-$, one finds out
that the
former ones prevail. The dominant ratios for the transitions $1^-\to 0^-$ are those corresponding to
even values for the angular momentum.

Now let us turn our attention to the electric transitions E1 and E3.
In Tables X and XI we listed the calculated E1 branching ratios:
\begin{eqnarray}
R^{10}_{+-}&=&\left[\frac{\langle \varphi^{(1,+)}_{J}||E_1||\varphi^{(g,-)}_{J+1}\rangle}
{\langle \varphi^{(1,+)}_{J}||E_1||\varphi^{(g,-)}_{J-1}\rangle}\right]^2,\nonumber\\
R^{10}_{-+}&=&\left[\frac{\langle \varphi^{(1,-)}_{J}||E_1||\varphi^{(g,+)}_{J+1}\rangle}
{\langle \varphi^{(1,-)}_{J}||E_1||\varphi^{(0,+)}_{J-1}\rangle}\right]^2.\nonumber\\
\end{eqnarray}
Concerning the ratio $R^{10}_{-+}$, two situations have been considered, namely when
the transition operator is harmonic and when the anharmonic term defined above has been included.
In that case we need the ratio $q_{anh}/q_h$. This ratios are to be fixed
so that a certain experimental data for the branching ratio is reproduced.
Such experimental data are available for the cases of $^{172}$Yb and $^{226}$Ra and
the determined
values for the ratio of anharmonic and harmonic weights of the transition operator
are equal to -1.722 and -1.4 respectively. The first value has been adopted also for
$^{158}$Gd while the second value was assigned for the ratio characterising the remaining nuclei
considered here.

\begin{table}
\begin{tabular}{|c|ccc|ccc|ccc|ccc|}
\hline
& \multicolumn{3}{c}{$^{158}$Gd}& \multicolumn{3}{c}{$^{172}$Yb}& \multicolumn{3}{c}{$^{228}$Th}& \multicolumn{3}{c|}{$^{232}$Th}\\
\hline
J & $1^+\to 0^-$ & $1^-\to 0^+$ &$1^-\to 0^+$ &$1^+\to 0^-$ & $1^-\to 0^+$ &$1^-\to 0^+$&$1^+\to 0^-$& $1^-\to 0^+$&$1^-\to 0^+$ & $1^+\to 0^-$& $1^-\to 0^+$ &$1^-\to 0^+$\\
\hline
1 &      &11.637 &5.321  &        &28.7    & 6.537 &       &13.197  & 5.991 &      &16.013&6.329 \\
2 &3.156 &       &       &2.641   &        &       & 3.065 &        &       &2.939 &   &  \\
3 &      & 1.948 &1.871  &        &2.056   &1.828  &       &1.955   &1.869  &      &1.974   & 1.858    \\
4 &1.559 &       &       &1.299   &        &       &1.512  &        &       &1.448&     &         \\
5 &      & 1.311 &1.598  &       &1.305   &1.478  &       &1.301   &1.546   &    &1.292    &1.516       \\
6 &1.650 &       &       &1.407  &        &       & 1.608 &        &        &1.550&     &         \\
7 &      &7.144  &2.899  &       & 8.280  &2.470  &       &7.163   & 2.967  &     &7.290    &2.849         \\
8 &2.221 &       &       &1.935  &        &       & 2.177 &        &        &2.112&     &           \\
9 &      & 7.312 &3.022  &       &7.197   &2.452  &       &7.157   &3.048   &     &7.025    &2.892          \\
10&5.365 &       &       &4.089  &        &       & 5.136 &       &         &4.825&     &            \\
11&      & 8.294 &3.252  &       &7.252   &2.539  &       &7.976  & 3.252   &      &7.624   &3.058             \\
12&5.909 &       &       &4.415  &        &       &5.627  &       &         &5.255 &    &            \\
13&      & 9.594 &3.520  &       &7.630   &2.666  &       &9.099  &3.499    &      &8.518   &3.266             \\
14&6.774 &       &       &4.899  &        &       &6.405  &       &         &5.928 &    &        \\
15&      &11.115 &3.808  &       &8.194   &2.816  &       &10.429 &3.768    &      &9.603    &3.498                \\
16&7.815 &       &       &5.453  &        &       &7.336  &        &        &6.726 &    &                    \\
17&      & 12.846&4.108  &       & 8.906  &2.982  &       &11.952  &4.052   &     &10.861    &3.746               \\
18&9.011 &       &       &6.066  &        &       &8.401  &       &         &7.632 &    &               \\
19&      &14.771 &4.417  &       &9.742   &3.158  &       &13.652 &4.346    &      &12.276    &4.005                \\
20&10.356&       &       &6.737  &        &       &9.597  &       &         &8.644 &    &         \\
21&      &16.877 &4.732  &       &10.687  &3.342  &       &15.517 &4.646    &      &13.836    &4.271         \\
22&11.846&       &       &7.469  &        &       &10.921 &       &         &9.762 &    &        \\
23&      &19.152 &5.050  &       &11.732  &3.532  &       &17.537 &4.951    &      &15.532    &4.542         \\
24&13.480&       &       &8.279  &        &       &12.373 &        &        &10.988&    &        \\
25&      &21.587 &5.370  &       &12.870  &3.727  &       &19.704  &5.258   &     &17.358    &4.816         \\
26&15.257&       &       &9.149  &        &       &13.952 &       &         &12.323&    &        \\
27&      &23.944 &5.686  &       &13.646  &3.915  &       &21.755  &5.560   &     &19.008    &5.085          \\
28&17.188&       &       &10.128 &        &       &15.676 &       &         &13.789&    &           \\
29&      &26.073 &6.013  &       &14.234  &4.131  &       &23.574 &5.877    &     &20.426   &5.374           \\
30&      &       &       &       &        &       &       &       &         &     &    &        \\
\hline
\end{tabular}
\caption{Calculated E1 branching ratios for the $K^{\pi}=1^+$ and $K^{\pi}=1^-$ bands
in the isotopes for $^{158}$Gd, $^{172}$Yb, $^{228}$Th, $^{232}$Th. Results given in the first and second columns
are obtained with a harmonic structure for the transition operator while those listed in the third
column correspond to an anharmonic structure given by Eq. (\ref{E1andM1}) with $q_{anh}/q_h$
equal to $-1.722$ for $^{158}$Gd and $^{172}$Yb and $-1.4$ for $Th$ isotopes.
These values were obtained by fitting the experimental value corresponding to
the state $1^-$ in
$^{172}$Yb  and $^{226}$Ra, respectively.}
\label{Table X}
\end{table}

\begin{table}[ht]
\begin{tabular}{|c|ccc|ccc|ccc|ccc|}
\hline
& \multicolumn{3}{c}{$^{226}$Ra}& \multicolumn{3}{c}{$^{238}$U}& \multicolumn{3}{c|}{$^{238}$Pu}\\
\hline
J & $1^+\to 0^-$ & $1^-\to 0^+$ &$1^-\to 0^+$ &$1^+\to 0^-$ & $1^-\to 0^+$ &$1^-\to 0^+$&$1^+\to 0^-$& $1^-\to 0^+$&$1^-\to 0^+$\\
\hline
1 &      &10.500 &5.815  &        & 37.24  &8.036  &       & 39.380 &8.072        \\
2 &3.141 &       &       &2.518   &        &       &2.521  &        &          \\
3 &      &1.89   &1.840  &        &2.090   &1.850  &       &2.108   &1.858           \\
4 &1.545 &       &       &1.236   &        &       &1.239  &        &           \\
5 &      &1.294  &1.523  &       & 1.321  &1.443   &       &1.325   &1.449             \\
6 &1.635 &       &       &1.344  &        &       &1.347  &        &         \\
7 &      &5.954  &2.932  &       & 8.531  &2.542  &       &9.178   &2.567               \\
8 &2.188 &       &       &1.845  &        &       &1.851  &        &              \\
9 &      &6.104  &2.988  &       & 7.178  &2.460  &       &7.569   &2.484              \\
10&5.100 &       &       &3.742  &        &       &3.778  &       &              \\
11&      &7.010  &3.156  &       & 7.087  &2.503  &       &7.349  &2.528            \\
12&5.624 &       &       &4.043  &        &       &4.072  &       &             \\
13&      &8.402  &3.364  &       &7.326   &2.591  &       &7.502  &2.619             \\
14&6.464 &       &       &4.474  &        &       &4.498  &       &          \\
15&      &9.341  &3.525  &       &7.631   &2.696  &       &7.858  &2.738        \\
16&7.491 &       &       &4.957  &        &       &4.976  &       &           \\
17&      &10.375 &3.688  &       &8.079   &2.817  &       &8.364  &2.875              \\
18&8.510 &       &       &5.467  &        &       &5.497  &       &           \\
19&      &11.460 &3.851  &       &8.632   &2.947  &       &8.989  &3.024              \\
20&9.628 &       &       &6.019  &        &       &6.062  &       &          \\
21&      &12.554 &4.015  &       &9.269   &3.085  &       &9.715  &3.182             \\
22&10.827&       &       &6.612  &        &       &6.670  &       &         \\
23&      &13.671 &4.182  &       &9.976   &3.226  &       &10.532 &3.347             \\
24&12.091&       &       &7.260  &        &       &7.339  &       &         \\
25&      &14.807 &4.356  &       &10.741  &3.369  &       &11.430 &3.518        \\
26&13.412&       &       &7.965  &        &       &8.069  &       &          \\
27&      &15.898 &4.533  &       &11.105  &3.500  &       &11.857 &3.676              \\
28&14.791&       &       &8.766  &        &       &8.906  &       &           \\
29&      &16.945 &4.718  &       &11.371  &3.657  &       &12.149 &3.875             \\
30&      &       &       &       &        &       &       &       &          \\
\hline
\end{tabular}
\caption{Calculated E1 branching ratios for the $K^{\pi}=1^+$ and $K^{\pi}=1^-$
bands,
in the isotopes  $^{226}$Ra, $^{238}$U, $^{238}$Pu. Results given in the first and
second columns
are obtained with a harmonic structure for the transition operator, while
those listed in the third
column correspond to an anharmonic structure given by Eq. (\ref{E1andM1}) with $q_{anh}/q_h=-1.4$.
This value was obtained by fitting the experimental value corresponding to
the state $1^-$ in
$^{226}$Ra.}
\label{Table XI}
\end{table}

In Fig. 10, the calculated branching ratios for the transitions of the negative
dipole band to the ground band are compared with the corresponding
data for the case of $^{172}$Yb. One notices  a good agreement between the two sets of data.
In Fig. 10 we also compare the calculated and experimental branching ratios
for the band $0^-$. One sees that these ratios are slowly decreasing with J up
to J=11 when a plateau is reached. By contrast, the branching for the dipole band is
decreasing up to $J=5$, has a small maximum at $J=7$, a flat minimum at J=9 and then is increasing with J.
Note that the dipole band has larger branchings than the band $0^-$. One notes that while the two experimental ratios for the band $0^-$ are quite well described by those predicted by the Alaga rule, i.e. 2.0 and 1.33 respectively large deviations for Alaga rule are  seen for the branching ratios characterising the dipole band with $K^{\pi}=1^-$. For example for J=1 and 3 the experimental ratios are about 6.5 and 2 respectively while the predictions of the Alaga rule are 0.5 and 0.75.
Our results and Alaga rule predictions are at variance not only in the region of low spin but also for high spin states.
Indeed, for J larger than nine, the Alaga rule predictions are close to 0.9 while in our case, starting with J=9 where the attained value is equal to about 2.5, our calculated ratios are increasing with J.

In Fig. 11 we compare three sets of data, namely the theoretical and experimental
branching ratios characterising the band $0^-$ and the branching ratio
associated to the negative parity dipole band, in $^{226}$Ra. For a better
representation, the
dipole branching ratios have been divided by 5. Remarkable the fact that
 modulo the factor five, the
calculated ratios for the two bands have similar behaviour as function of J.
It is interesting to see how the matrix elements of the E1 transition operator
depend on the angular momentum,
for the bands with $K^{\pi}=0^-$ and $K^{\pi}=1^-$.
Comparison of the two sets of matrix elements is made in Fig.12.
Note that for $J=odd$, the matrix element is characterising the transition from
 the states $J^-$ to the state $(J-1)^+$
while for $J=even$, this links the states $J^+$ and $(J-1)^-$. For both
situations an
anharmonic structure for the transition operator has been considered.
The  parameters  involved in the $T_{1\mu}$ operator have the values:
$q_{anh}/q_h=-1.4$ and $q_h=10^{-2} fm$.
We note that the matrix elements describing the transition of the dipole state
$J^-$, multiplied by a factor of 5 stays quite close to the similar matrix elements characterising the band $0^-$.

In Fig. 13 we study  the intraband E2 transitions in the two dipole bands. The calculated
 B(E2) values are divided by the B(E2) value corresponding to the transition
 $2^{\dagger}_g\to 0^{\dagger}_g$.
\begin{equation}
R_q=\frac{B(E2;J\to (J-2))}{B(E2;2^+_g\to 0^+_g)}.
\label{ErQ}
\end{equation}

Results were obtained with a harmonic quadrupole
transition operator and by neglecting the admixture of gamma band states in the structure of the
ground band states. For comparison, we plotted also the reduced transition probability
in the ground band, with a similar normalisation. The intraband quadrupole transitions have similar dependence on angular momentum in the two dipole bands.
Moreover, the B(E2) values for the transitions $J^-\to (J-2)^-$ with J even lie
close to the curve corresponding to the ground band.

The dipole bands may perform a E3 transition to the ground bands. To study the
E3 properties of the dipole bands we have used an harmonic transition operator:

\begin{equation}
T_{3\mu}=q_3\left(b^{\dagger}_{3\mu}+(-)^{\mu}b_{3,-\mu}\right).
\end{equation}
These transitions turns out to be weak. To have a flavour about the
relative values of these transitions,
comparing them to those from the band $0^-$, we normalise each transition to
the B(E3) value associated for the transition $3^-_g\to 0^+_g$.

In Table XII we list the calculated values for the ratios:
\begin{eqnarray}
T^{10}_{-+}(3;J)&=&\frac{B(E3,J^-\to (J+3)^+_g)}{B(E3;3^-_g\to 0^+_g)},\nonumber\\
T^{10}_{+-}(3;J)&=&\frac{B(E3,J^+\to (J+3)^-_g)}{B(E3;3^-_g\to 0^+_g)}.
\end{eqnarray}

\begin{table}[ht]
\begin{tabular}{|ccc|}
\hline
J  & $\hskip0.5cm$$T^{10}_{-+}(3;J)$$\hskip0.5cm$ & $\hskip0.5cm$ $T^{10}_{+-}(3;J)$$\hskip0.5cm$ \\
\hline
1   &0.002&     \\
2   &     &0.0  \\
3   &0.012&     \\
4   &     &0.011\\
5   &0.030&     \\
6   &     &0.021\\
7   &0.057&     \\
8   &     &0.036\\
9   &0.090&      \\
10  &     &0.054 \\
11  &0.126&      \\
12  &     &0.076 \\
13  &0.163&      \\
14  &     &0.102 \\
15  &0.199&      \\
16  &     &0.133 \\
17  &0.234&      \\
18  &     &0.167 \\
19  &0.267&      \\
20  &     &0.205 \\
21  &0.297&      \\
22  &     &0.244 \\
23  &0.328&      \\
24  &     &0.284 \\
25  &0.358&      \\
26  &     &0.325 \\
27  &0.389&      \\
\hline
\end{tabular}
\caption{The calculated values for the B(E3) values, normalized to the B(E3)
value for the transition
$3^-_g\to 0^+_g$, are listed for the transitions of the dipole band
states of angular momentum J to the states  of angular momentum (J+3) from the ground bands.
The calculations were performed for $^{226}$Ra.}
\label{Table XII}
\end{table}

\section{Conclusions}
\renewcommand{\theequation}{5.\arabic{equation}}
\setcounter{equation}{0}

\label{sec:level5}

In the previous sections we developed a formalism which appends the description of dipole bands to the extended
coherent state model which results in obtaining a simultaneous and consistent model     of
eight rotational bands, four of positive and four of negative parity. Since
for the seven nuclei
considered, the results for three parity partner bands were already reported
in some earlier publications
here we focus on the description of the dipole bands. The present
paper is the first which  is
devoted to the formalism description giving the analytical results describing the
states and the matrix elements of the model Hamiltonian and transition operators.

The eight rotational bands are obtained by projecting out the angular
momentum and parity, from four intrinsic states which are quadrupole and octupole deformed functions
and moreover orthogonal onto each other. By construction the four states have
the property that the
eight sets of projected states are all orthogonal. The model states depend on two real parameters which simulate
the quadrupole and octupole deformation, respectively. In the spherical limit i.e.,
both deformations tend to zero,
specific multiphonon states of definite angular momentum, seniority and
number of bosons respectively,  are obtained. In the large deformation regime the projected
states have a definite value for the  $K$ quantum number.
In the restricted space of projected states, an effective quadrupole-octupole boson Hamiltonian is considered.
Indeed, for the model Hamiltonian the only nonvanishing matrix elements are those involving the states $(J^{\dagger}_g,J^{\dagger}_{\gamma})$ with $J=even$ and
$(J^{-}_g,J^{-}_{\gamma})$ with $J=odd$. The structure coefficients defining the model
Hamiltonian as well as the deformation parameters have been fixed by fitting
through the least square procedure the energies in the bands $g^{\pm},\beta^{\pm},\gamma^{\pm}$
and the energy of the head state in the $K^{\pi}=1^-$ band. Also, one parameter (${\cal C}_2$ has been determined so that
the contribution of the ${\cal B}_3$ term to a particular state ($2^{-}$) is canceled.
This condition seems to be sufficient to decrease the
off diagonal matrix elements involving the dipole band states  to
negligible values.

It is worth noting that dynamic moment of inertia of the odd and even angular momenta states
are lying on separate smooth curves which could suggest that the two sets of states form distinct
bands. This happens for both the positive and negative dipole bands. However,
two pairs  of these curves one for positive and one for negative subsets of states
have an interleaved structure. These made us suspecting that an octupole
static deformation shows up. In order to get a confirmation for this 
suspicion
we calculated the first order and the second order energy displacement functions.
Both functions vanish for similar angular momenta in $^{172}$Yb, $^{226}$Ra, $^{238}$U and
$^{238}$Pu. The only isotope where the vanishing persists in a relatively long range of angular momentum,
is $^{228}$Ra. For other nuclei mentioned above the vanishing takes place in 1-3 states.
In Ref.\cite{Rad31} we interpreted the vanishing of the displacement function for a
very short interval of J
in a way which conciliate between the band intersection and static octupole deformation.
Indeed, for such states it may happen that they could be obtained by projection
from an octupole deformed  state which is different from the chosen model state for the dipole bands.
Moreover, from this deformed state one could generate, through the angular momentum projection procedure,
another two bands which are deformed all along and intersect the dipole bands
considered here for the mentioned angular momentum. In this respect one could assert that bands intersection does not exclude
the octupole deformation settlement.

Since the decay properties of the states depend on the corresponding
boson structure, we calculated the angle between the angular momenta carried by the two kind of bosons.
The result is that for high angular momentum states, this angle approaches the value
$\pi/2$.  This value is reached first in the negative and then in the positive band.
Exception is for $^{226}$Ra and $^{238}$U where the angles in the two bands 
go
simultaneously to the limit value $\pi/2$. We expect that for these systems, by adding a coupling
set of particles one could  reach a chirally symmetric picture.

For the sake of a complete description of the dipole states of positive and
negative parity, by using
the eigenstates of the model Hamiltonian,  the intra and interband transitions
of electric as well as of magnetic nature have been calculated.
Comparison with the experimental data is made in terms of the branching ratios of the negative parity states.
Also, these are compared with those characterizing the negative parity band with
$K^{\pi}=0^-$.
Also the gyromagnetic factors for the two bands were calculated. One notices a
strong dependence on angular momentum for the gyromagnetic factors.

Comparing the intraband B(M1) values obtained for the two dipole bands, one concludes that
the strength of magnetic transitions in the band $K^{\pi}=1^+$ is larger than that one associated with the
band $K^{\pi}=1^-$. Due to this feature we say that the $K^{\pi}=1^+$ band has a magnetic nature.
Concerning the $E1$ transition to the parity partner band of to the $g^{\pm}$ band, the
strength order is changed. Therefore we say that the band $K^{\pi}=1^-$ has an electric nature.
It is worth  mentioning the role of parity projection in determining the
magnetic or electric nature of the two bands.

Now let us  say few words about the distinctive features of our formalism.
The procedure is interesting not only because is able to describe a relatively large
volume of data with a relatively small number of parameters, but also because it
provides a consistent description of the rotational degrees of freedom.
Indeed, all formalisms based on quadrupole and octupole boson interaction
overestimate the contribution of the rotational degrees of freedom. That happens since in the
intrinsic frame the Eulerian angles associated to the quadrupole and octupole coordinates
are independent variables. Such a redundancy is automatically removed in the
present formalism due to the projection operation. Another salient feature of
the coherent state formalism consists of that it represents the ideal framework
for the description of the semiclassical aspects of the collective motion.
In particular, it provides a suitable description for the high spin states,
where the nuclear system behaves semiclassically, as well as for the
quadrupole and octupole deformed systems.

Moreover, the mechanism for a static
octupole deformation
is different \cite{Rad12} from the traditional one where a fourth order octupole boson term is
necessary \cite{Jol}. As explained in  Ref.\cite{Rad12}, in our formalism a second order octupole boson term is sufficient for obtaining
a stable octupole deformed shape.

An octupole shaped system may have nonvanishing electric dipole moment.
 Also, due to the fact that the angular momentum is built up by
both quadrupole and octupole bosons, one expects that the magnetic properties in a given state
depend on its boson composition. Such properties may show up
in dipole bands. Up to now, the set on of the octupole deformation was
associated with
a jump in the dipole matrix element $J^-\to (J-1)^+_g$ (see the case of $^{226}$Ra), where $J^-$ belongs to  the band $0^-$.
We pointed out that the positive parity state
having static octupole deformation (the value of J where the energy displacement function vanishes)
exhibits large M1 branching ratio to the ground band.

Before closing this section, we want to comment on the nature of the excited bands.
Many authors believe that the
states of non-vanishing
$K$ cannot be of collective nature. To give an example,
the authors of Ref.\cite{Wal} invoke the arguments from Ref. \cite{Vog} and
interpret the dipole states
of negative parity in $^{172}$Yb, as two quasi-neutron states.
On the other hand,   based on microscopic studies with surface delta
interaction,  the authors of Ref. \cite{Fas}, concluded that the
$K^{\pi}=1^-,2^-$ bands of some actinides have, however, a collective nature.
Actually, this is not the only example in the literature when one proves that
 the  microscopic interpretation of the negative parity states,
as two or four quasiparticle states is not unique. Indeed, the
double bending, one back and one forward,  seen
in the ground and $0^-$ bands of $^{218}$Ra, interpreted in Ref.\cite{Schu} as
caused by  successive intersections of a collective band, a two neutron and
a two neutron plus two proton quasiparticle bands, are fairly well reproduced
by the phenomenological description provided by ECSM \cite{Rad110}.
Although  the dipole states for $^{172}$Yb are considered in Ref. \cite{Wal}
as two quasi-neutron states, the branching ratios of the $K^{\pi}=0^-,1^-$
low lying states are
realistically described within an IBA-sdf formalism  in Ref.\cite{Bret}. 
Moreover, as we have already shown, the
present paper provides also a good description of the electric transitions in this nucleus.
In the examples mentioned above the effect of single particle degrees of freedom
is simulated by the competition
between various anharmonic terms involved in the model Hamiltonian or in the
transition operator. More experimental data regarding both the excitation energies and transition probabilities in the dipole bands, would be a decisive test for the predictable power of our formalism.

\section{Appendix A}
\renewcommand{\theequation}{A.\arabic{equation}}
\setcounter{equation}{0}
\label{sec:levelA}

Here we shall list the explicit expressions for the norms of all projected states
defined in the previous sections.
The norm of the states obtained by projecting out the angular momentum and the parity
from the octupole boson coherent state, has the expression:
\begin{equation}
\left(N^{(\pm)}_{oc,J}\right)^{-2}=e^{-y_3}(2J+1){\cal I}^{(\pm)}_J(y_3),~y_3=f^2,
\end{equation}
where ${\cal I}^{(\pm)}$ stands  for the overlap functions:
\begin{eqnarray}
{\cal I}^{(+)}_J(y_3)&=&\int_0^1 P_J(x)ch\left[f^2P_3(x)\right]dx,\nonumber\\
{\cal I}^{(-)}_J(y_3)&=&\int_0^1 P_J(x)sh\left[f^2P_3(x)\right]dx,
\end{eqnarray}
with $P_J(x)$ denoting the Legendre polynomial of rank J.

The norms of the dipole states are expressed in terms of norms characterising the
projected states associated to the quadrupole and octupole state factors:

\begin{eqnarray}
\left(N^{(1,\pm)}_J\right)^{-2}&=&\sum_{J_2,J_3}\left(N^{(\pm)}_{31;J_3}\right)^{-2}
\left(N^{(g)}_{J_2}\right)^{-2}\left(C^{J_3\;J_2\;J}_{1\;\;0\;\;1}\right)^2,
\nonumber\\
\left(N^{(\pm)}_{31;JM}\right)^{-2}&=&\frac{1}{{\cal A}}
C^{J'\;1\;J}_{0\;\;1\;\;1}\sqrt{J(J+1)}\left(N^{(\pm)}_{3,J}\right)^{-1},
\nonumber\\
\left(N^{(\pm)}_{3,J}\right)^{-2}&=&\left(N^{(\pm)}_{oc,J}\right)^{-2}\left(2+\frac{4}{7}f^2
\frac{{\cal I}^{(\pm)\prime}_J}{{\cal I}^{(\pm)}_J} \right),
\nonumber\\
{\cal A}&=&-\sqrt{12}C^{3\;\;3\;\;1}_{1\;\;0\;\;1}f.
\end{eqnarray}
The standard notation for the Clebsch-Gordan coefficient $C^{j_1 J_2 j}_{m_1 m_2 m}$ has been used.
\section{Appendix B}
\renewcommand{\theequation}{B.\arabic{equation}}
\setcounter{equation}{0}
\label{sec:levelB}

Here we give the analytical expressions for the matrix elements of the terms
involved in the model Hamiltonian. For what follows it is useful to introduce the notation:
\begin{equation}
^{(1)}X^{J_2J_3}_{Jk}=\left(N^{(1,k)}_J\right)^2\left(N^{(k)}_{31;J_3}N^{(g)}_{J_2}\right)^{-2}
\left(C^{J_3\;J_2\;J}_{1\;\;0\;\;1}\right)^2.
\end{equation}
The final results for the matrix elements are:

\begin{eqnarray}
\langle \varphi^{(1,\pm)}_{JM}|{\hat N}_2|\varphi^{(1,\pm)}_{JM}\rangle & =&
\sum_{J_2,J_3}{^{(1)}X^{J_2J_3}_{J\pm}}d^2\frac{I^{(1)}_{J_2}}{I^{(0)}_{J_2}},
\nonumber\\
\langle \varphi^{(1,\pm)}_{JM}|{\hat N}_3|\varphi^{(1,\pm)}_{JM}\rangle &=&
\sum_{J_2,J_3}
\left[2+\frac{18}{7}f^2\frac{{\cal I}^{(\pm)'}_{J_3}}
{{\cal I}^{(\pm)}_{J_3}} \left(N^{(\pm)}_{oc;J_3}\right)^{-2}
+\frac{4}{7}f^4\sum_{J_1}\left(C^{J_1\;3\;J_3}_{0\;\;0\;\;0}\right)^2
\frac{{\cal I}^{(\mp)^{\prime}}_{J_1}}
{{\cal I}^{(\mp)}_{J_1}}\left(N^{(\mp)}_{oc;J_1}\right)^{-2}\right],
\nonumber\\
&\times&{^{(1)}X^{J_2J_3}_{J\pm}}\left(N^{(\pm)}_{3;J_3}\right)^2
\nonumber\\
\langle \varphi^{(1,\pm)}_{JM}|{\hat N}_2 {\hat N}_3 |\varphi^{(1,\pm)}_{JM}\rangle &=&
\sum_{J_2,J_3}
\left[2+\frac{18}{7}f^2\frac{{\cal I}^{(\pm)'}_{J_3}}
{{\cal I}^{(\pm)}_{J_3}}\left(N^{(\pm)}_{3;J_3}\right)^2  \left(N^{(\pm)}_{oc;J_3}\right)^{-2}
+\frac{4}{7}f^4\left(C^{J_1\;3\;J_3}_{0\;\;0\;\;0}\right)^2
\frac{{\cal I}^{(\mp)'}_{J_1}}
{{\cal I}^{(\mp)}_{J_1}}\left(N^{(\pm)}_{3;J_3}\right)^2  \left(N^{(\mp)}_{oc;J_1}\right)^{-2}\right]
\nonumber\\
&\times& {^{(1)}X^{J_2J_3}_{J\pm}}d^2\frac{I^{(1)}_{J_2}}{I^{(0)}_{J_2}},\nonumber\\
\langle \varphi^{(1,\pm)}_{JM}|\vec{J}_2 \vec{J}_3 |\varphi^{(1,\pm)}_{JM}\rangle &=&
\sum_{J_2,J_3}{^{(1)}X^{J_2J_3}_{J\pm}}
\left[J(J+1)-J_2(J_2+1)-J_3(J_3+1)\right],
\end{eqnarray}
\begin{eqnarray}
\langle \varphi^{(1,\pm)}_{JM}|\Omega^{\dag}\Omega |\varphi^{(1,\pm)}_{JM} \rangle &=&
\sum_{J_2J_3}   \left[4+\frac{144}{49}f^2
\frac{{\cal I}^{(\pm)\prime}_{J_3}}{{\cal I}^{(\pm)}_{J_3}}
\left(N^{(\pm)}_{oc,J_3}\right)^{-2} +
\frac{16}{49}f^4\sum_{J^{\prime}_3}\left(C^{J^{\prime}_3\;3\;J_3}_{0\;;0\;;0}\
\right)^2
\frac{{\cal I}^{(\mp)\prime}_{J^{\prime}_3}}{{\cal I}^{(\mp)}_{J^{\prime}_3}}
\left(N^{(\mp)}_{oc,J^{\prime}_3}\right)^{-2} \right]
\nonumber\\
{^{(1)}X^{J_2J_3}_{J\pm}}\left(N^{(\pm)}_{3J_3}\right)^2,
\nonumber\\
\langle \varphi^{(1,\pm)}_{JM}|\Omega^{\dag}\hat{N}_2\Omega |\varphi^{(1,\pm)}_{JM} \rangle &=&
\sum_{J_2J_3}   \left[4+\frac{144}{49}f^2
\frac{{\cal I}^{(\pm)\prime}_{J_3}}{{\cal I}^{(\pm)}_{J_3}}\left(N^{(\pm)}_{oc,J_3}\right)^{-2} +
\frac{16}{49}f^4\sum_{J^{\prime}_3}\left(C^{J^{\prime}_3\;3\;J_3}_{0\;;0\;;0}\right)^2
\frac{{\cal I}^{(\mp)\prime}_{J^{\prime}_3}}{{\cal I}^{(\mp)}_{J^{\prime}_3}}
\left(N^{(\mp)}_{oc,J^{\prime}_3}\right)^{-2} \right]\nonumber\\
&\times&{^{(1)}X^{J_2J_3}_{J\pm}}\left(N^{(\pm)}_{3J_3}\right)^2
d^2\frac{I^{(1)}_{J_2}}{ I^{(0)}_{J_2}}.
\end{eqnarray}

\section{Appendix C}
\renewcommand{\theequation}{C.\arabic{equation}}
\setcounter{equation}{0}
\label{sec:levelC}

The reduced matrix elements for the harmonic part of the E1 transition operator
relating the dipole states to the states from the ground bands are:

\begin{eqnarray}
\langle \varphi^{(1,+)}_{J}\parallel T^{(h)}_1\parallel \varphi^{(g,-)}_{J^{\prime}}\rangle
&=&q_1N^{(1,+)}_JN^{(g,-)}_{J^{\prime}}
\sum_{J_2,J_3,J^{\prime}_2,J^{\prime}_3}C^{J_3\;J_2\;J}_{1\;\;0\;\;0\;\;1}C^{J^{\prime}_3\;J^{\prime}_2\;J^{\prime}}_{0\;\;0\;\;0}
\left(N^{(+)}_{31J_3}\right)^{-1} \left(N^{(g,-)}_{J^{\prime}}\right)^{-1}
\nonumber\\
&\times&
\langle\varphi^{(+)}_{3,J_3} \parallel b^{\dag}_3+b_3\parallel\varphi^{(-)}_{oc;J^{\prime}_3}\rangle
\langle\varphi^{(g)}_{J_2} \parallel b^{\dag}_2+b_2\parallel\varphi^{(g)}_{J^{\prime}_2}\rangle,
\nonumber\\
\langle \varphi^{(1,-)}_{J}\parallel T^{(h)}_1\parallel \varphi^{(g,+)}_{J^{\prime}}\rangle
&=&q_1N^{(1,-)}_JN^{(g,+)}_{J^{\prime}}
\sum_{J_2,J_3,J^{\prime}_2,J^{\prime}_3}C^{J_3\;J_2\;J}_{1\;\;0\;\;0\;\;1}C^{J^{\prime}_3\;J^{\prime}_2\;J^{\prime}}_{0\;\;0\;\;0}
\left(N^{(-)}_{31J_3}\right)^{-1} \left(N^{(g,+)}_{J^{\prime}}\right)^{-1}
\nonumber\\
&\times&
\langle\varphi^{(-)}_{3,J_3} \parallel b^{\dag}_3+b_3\parallel\varphi^{(+)}_{oc;J^{\prime}_3}\rangle
\langle\varphi^{(g)}_{J_2} \parallel b^{\dag}_2+b_2\parallel\varphi^{(g)}_{J^{\prime}_2}\rangle,
\nonumber\\
\langle\varphi^{(+)}_{3,J_3} \parallel b^{\dag}_3+b_3\parallel\varphi^{(-)}_{oc;J^{\prime}_3}\rangle
&=&\frac{2}{\sqrt{7}}\frac{\hat{J}^{\prime}_3}{\hat{J}_3}f
N^{(+)}_{3;J_3}N^{(-)}_{oc;J^{\prime}_3}\left(N^{(-)}_{oc;J_3}\right)^{-2}
C^{J_3\;3\;J^{\prime}_3}_{0\;\;0\;\;0},
\nonumber\\
\langle\varphi^{(-)}_{3,J_3} \parallel b^{\dag}_3+b_3\parallel\varphi^{(+)}_{oc;J^{\prime}_3}\rangle
&=&\frac{2}{\sqrt{7}}\frac{\hat{J}^{\prime}_3}{\hat{J}_3}f
N^{(-)}_{3;J_3}N^{(+)}_{oc;J^{\prime}_3}\left(N^{(+)}_{oc;J_3}\right)^{-2}
C^{J_3\;3\;J^{\prime}_3}_{0\;\;0\;\;0}.
\end{eqnarray}

The transition operator involves an anharmonic term, $T^{anh}_{1\mu}$.
Due to this component of the transition operator a given state from a dipole band can decay to a state
from the ground band of opposite parity:
\begin{eqnarray}
\langle \varphi^{(1,\pm)}_J\parallel T^{anh}_1\parallel\varphi^{(g,\mp)}_{J^{\prime}} \rangle &=&
N^{(1,\pm)}_JN^{(g,\mp)}_{J^{\prime}}\left(N^{(\pm)}_{31;J_3}\right)^{-1}\left(N^{(g)}_{J_2}\right)^{-2}
C^{J_3\;J_2\;J}_{1\;\;0\;\;1} C^{J^{\prime}_3\;J_2\;J^{\prime}}{0\;\;0\;\;0}
\nonumber\\
&\times &5\sqrt{15J_2\left(J_2+1\right)
J^{\prime}_3\left(J^{\prime}_3+1\right)}\hat{J}_2 \hat{J}_3 \hat{J}^{\prime}_3 \hat{J}^{\prime}
W\left(1113;22\right)W\left(J^{\prime}_31J_33;J^{\prime}_32\right)\nonumber\\
&\times&\sum_{J_4}(2J_4+1)W\left(J_3J_2J1;J_4J_2\right)W\left(J_3J_22J^{\prime};J_4J^{\prime}_3\right)
W\left(j^{\prime}2J1;J_41\right)\langle\varphi^{(\pm)}_{31;J_3}\parallel b^{\dag}_3
\parallel \varphi^{(\mp)}_{oc;J^{\prime}_3}\rangle,
\nonumber\\
\langle \varphi^{(+)}_{3;J_3}\parallel b^{\dag}_3+b_3\parallel \varphi^{(-)}_{oc;J^{\prime}_3}\rangle &=&
\frac{2\hat{J}^{\prime}_3f}{\sqrt{7}\hat{J}_3}N^{(+)}_{J_3}N^{(-)}_{oc;J^{\prime}_3}
\left(N^{(+)}_{oc;J^{_3}} \right)^{-2}C^{J_3\;3\;J^{\prime}_3}_{0\;\;0\;\;0}.
\end{eqnarray}

Taking for the E2 transition operator an harmonic form, the matrix elements describing
the transitions within the dipole bands are:
\begin{eqnarray}
\langle\varphi^{1,\pm}\parallel b^{\dag}_2+b_2\parallel\varphi^{(1,\pm)}_{J^{\prime}}\rangle &=&
dN^{(1,\pm)}_J N^{(1,\pm)}_{J^{\prime}}\hat{J}^{\prime}
\sum_{J_2J^{\prime}_2J_3}C^{J_2\;J_3\;J}_{0\;\;1\;\;1}
C^{J^{\prime}_2\;J_3\;J^{\prime}}_{0\;\; 1\;\;1}
C^{J^{\prime}_2\;2\;J_2}_{0\;\; 0\;\;0} \hat{J}_2W\left(2J_2J^{\prime}J_3;J^{\prime}_2J\right)
\nonumber\\
&\times&\left(N^{(\pm)}_{31;J_3}\right)^{-1}
\left(\left(N^{(g)}_{J^{\prime}_2}\right)^{-2}+\frac{2J^{\prime}_2+1}{2J_2+1}
\left(N^{(g)}_{J_2}\right)^{-2}\right).
\end{eqnarray}
The E3 operator
\begin{equation}
T_{3\mu}=q_3\left(b^{\dag}_{3\mu}+(-)^{1+\mu}b_{3-\mu}\right),
\end{equation}
relates a dipole state with a state from the corresponding ground band.
\begin{eqnarray}
\langle \varphi^{(1,\mp)}_J\parallel T_3\parallel \varphi^{(g,\pm)}_{J^{\prime}}\rangle &=&
q_3\frac{-2f}{\sqrt{7}}N^{(1,\mp)}_JN^{(g,\pm)}_{J^{\prime}}
\sum_{J_2,J_3,J^{\prime}_3}\hat{J}^{\prime}_3\hat{J}^{\prime}
C^{J_2\;J_3\;J}_{0\;\;1\;\;1} C^{J_2\;J^{\prime}_3\;J^{\prime}}_{0\;\;0\;\;0}
C^{J_3\;3\;J^{\prime}_3}_{0\;\;0\;\;0}
W\left(JJ_23J^{\prime}_3;J_3J^{\prime}\right)\nonumber\\
& \times &\left(N^{(\mp)}_{31;J_3}\right)^{-1}\left(N^{(g)}_{J_2}\right)^{-2}
N^{(\mp)}_{3;J_3}\left(N^{(\mp)}_{oc;J_3}\right)^{-2}.
\end{eqnarray}
The corresponding transition rate is compared with the octupole
strength  characterising the transition  $\varphi^{g,-}\to\varphi^{g,+}$.
\begin{eqnarray}
\langle \varphi^{(g,-)}_J\parallel b^{\dag}_3+b_3\parallel\varphi^{(g,+)}_{J^{\prime}}\rangle
&=&f N^{(g,-)}_JN^{(g,+)}_{J^{\prime}}
\sum_{J_2,J_3,J^{\prime}_3}C^{J_2\;J_3\;J}_{0\;\;0\;\;0}
C^{J_2\;J^{\prime}_3\;J^{\prime}}_{0\;\;0\;\;0} C^{J^{\prime}_3\;3\;J_3}_{0\;\;0\;\;0}
\hat{J}_3\hat{J}^{\prime}W\left(3J^{\prime}_3JJ_2;J_3J^{\prime}\right)
\nonumber\\
&\times&\left(N^{(g)}_{J_2}\right)^{-2}
\left(\left(N^{(+)}_{oc;J^{\prime}_3}\right)^{-2}+\frac{2J^{\prime}_3+1}{2J_3+1}
\left(N^{(-)}_{oc;J_3}\right)^{-2}\right).
\end{eqnarray}
The dipole states may decay to the ground band states of similar parity, 
by means of the
M1 transition operator defined by Eq.(4.4).

The nonvanishing matrix elements relating the dipole and ground band states are:

\begin{eqnarray}
\langle \varphi^{(1,+)}_J\parallel M_1\parallel\varphi^{g,+)}_{J^{\prime}}\rangle &=&
g^{\prime}_2N^{(1,+)}_JN^{(g,+)}_{J^{\prime}}\sum
C^{J_2\;J_3\;J}_{0\;\;1\;\;1}C^{J_2\;J^{\prime}_3\;J^{\prime}}_{0\;\;0\;\;0}
\left(N^{(+)}_{31;J_3}\right)^{-1} \left(N^{(g)}_{J_2}\right)^{-1}
\left(N^{(+)}_{oc;J^{\prime}_3}\right)^{-1} \sqrt{J_2(J_2+1)}\nonumber\\
&\times &T^{J_2J_3}_{JJ^{\prime}}
\langle \varphi^{(+)}_{3J_3}\parallel \left(b^{\dag}_3b^{\dag}_3\right)_2\parallel \varphi^{(+)}_{oc;J^{\prime}_3}\rangle ,
\nonumber\\
\langle \varphi^{(1,-)}_J\parallel M_1\parallel\varphi^{g,-)}_{J^{\prime}}\rangle &=&
g^{\prime}_2N^{(1,-)}_JN^{(g,-)}_{J^{\prime}}\sum
C^{J_2\;J_3\;J}_{0\;\;1\;\;1}C^{J_2\;J^{\prime}_3\;J^{\prime}}_{0\;\;0\;\;0}
\left(N^{(-)}_{31;J_3}\right)^{-1} \left(N^{(g)}_{J_2}\right)^{-1}
\left(N^{(-)}_{oc;J^{\prime}_3}\right)^{-1} \sqrt{J_2(J_2+1)}\nonumber\\
&\times &T^{J_2J_3}_{JJ^{\prime}}
\langle \varphi^{(-)}_{3J_3}\parallel \left(b^{\dag}_3b^{\dag}_3\right)_2\parallel \varphi^{(-)}_{oc;J^{\prime}_3}\rangle ,
\end{eqnarray}
where
\begin{eqnarray}
T^{J_2J_3}_{JJ^{\prime}}&=&\hat{1}\hat{J}_2\hat{J}_3\hat{J}^{\prime}\sum_{J_4}
(2J_4+1)W\left(J^{\prime}_3J_311;2J_4\right)W\left(J_2J^{\prime}_3J1;J^{\prime}J_4\right)
W\left(J_21JJ_3;J_2J_4\right),\nonumber\\
\langle \varphi^{(\pm)}_{3J_3}\parallel \left(b^{\dag}_3b^{\dag}_3\right)_2\parallel \varphi^{(\pm)}_{oc;J^{\prime}_3}\rangle
&=&\frac{4}{7}f^2\hat{2}\hat{J}^{\prime}_3N^{(\pm)}_{3J_3}N^{(\pm)}_{oc;J^{\prime}_3}
\sum_{J_1=odd}C^{J_1\;3\;J_3}_{0\;\;0\;\;0}C^{J_1 \;3\;J^{\prime}_3\;J^{\prime}}_{0\;\;\;0\;\;\;0}
W\left(J^{\prime}_32J_13;J_33\right)\left(N^{(\pm)}_{oc;J_1}\right)^{-2}.
\end{eqnarray}
As usual the abbreviation $\hat{J}=\sqrt{2J+1}$ is used. When one deals with the angular momentum operator, the symbol `` hat'' suggests the operatorial character.

The M1 transitions within the dipole bands as well as the gyromagnetic factors of the
dipole states were determined by restricting the transition operator to the lowest order boson terms:
\begin{equation}
M_{1\mu}=g_2\left(\hat{J}_{2}\right)_{\mu}+g_3\left(\hat{J}_{3}\right)_{\mu}.
\end{equation}
The transition amplitudes are given by the reduced matrix elements:
\begin{eqnarray}
\langle \varphi^{(1,\pm)}_J\parallel g_2J_{2}+g_3J_{3}\parallel\varphi^{(1,\pm)}_{J^{\prime}}\rangle &=&
N^{(1,\pm)}_J N^{(1,\pm)}_{J^{\prime}}
\sum_{J_2J_3}C^{J_2\;J_3\;J}_{0\;\;1\;\;1}C^{J_2\;J_3\;J^{\prime}}_{0\;\;1\;\;1}\left(N^{(+)}_{31;J_3}\right)^{-2}
\left(N^{(g)}_{J_2}\right)^{-2}\\
& \times &\left[\hat{J}_2\hat{J}^{\prime}W\left(1J_2J^{\prime}J_3;J_2J\right)\sqrt{J_2(J_2+1)}g_2+
\hat{J}_3\hat{J}^{\prime}W\left(1J_3JJ_2;J_3J^{\prime}\right)\sqrt{J_3(J_3+1)}g_3\right].
\nonumber
\end{eqnarray}
Using these expressions one could calculate the gyromagnetic factors of the dipole states:
\begin{equation}
g^{\pm}_J=\frac{1}{\sqrt{J(J+1)}}\langle \varphi^{(1,\pm)}_J\parallel g_2\hat{J}_{2}+g_3\hat{J}_{3}\parallel\varphi^{(1,\pm)}_J\rangle .
\end{equation}

\newpage


\begin{figure}[ht]
\begin{center}
\includegraphics[height=10cm,width=8cm]{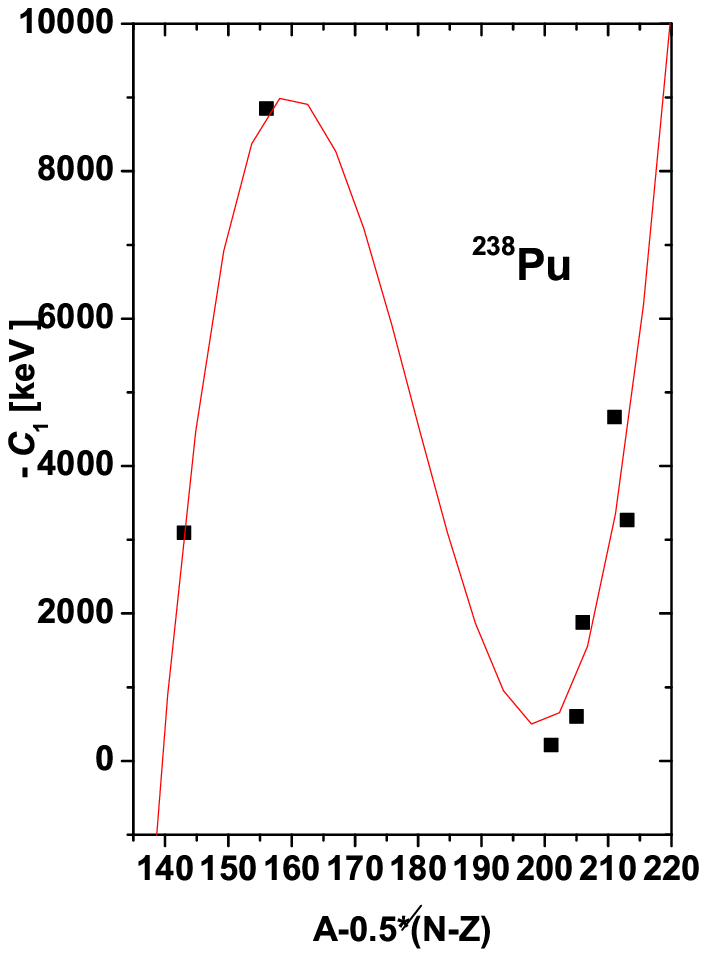}
\end{center}
\caption{The structure coefficient ${\cal C}_1$, determined as explained in the text, is represented as function
of $A-0.5*(N-Z)$ (black square). The obtained values are interpolated by a third order polynomial (full line curve).}
\label{Fig. 1}
\end{figure}
\clearpage

\begin{figure}[ht]
\begin{center}
\includegraphics[height=10cm,width=8cm]{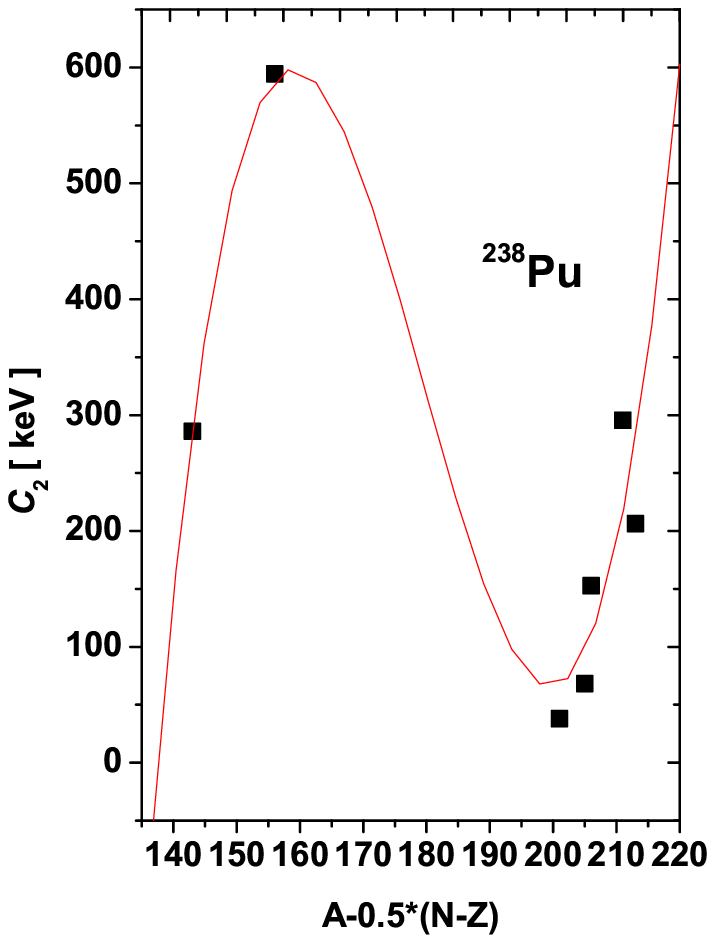}
\end{center}
\caption{The structure coefficient ${\cal C}_2$, determined as explained in the text, is represented as function
of $A-0.5*(N-Z)$ (black square). The obtained values are interpolated by a third order polynomial (full line curve).}
\label{Fig. 2}
\end{figure}
\clearpage

\begin{figure}[ht]
\begin{center}
\includegraphics[width=0.8\textwidth]{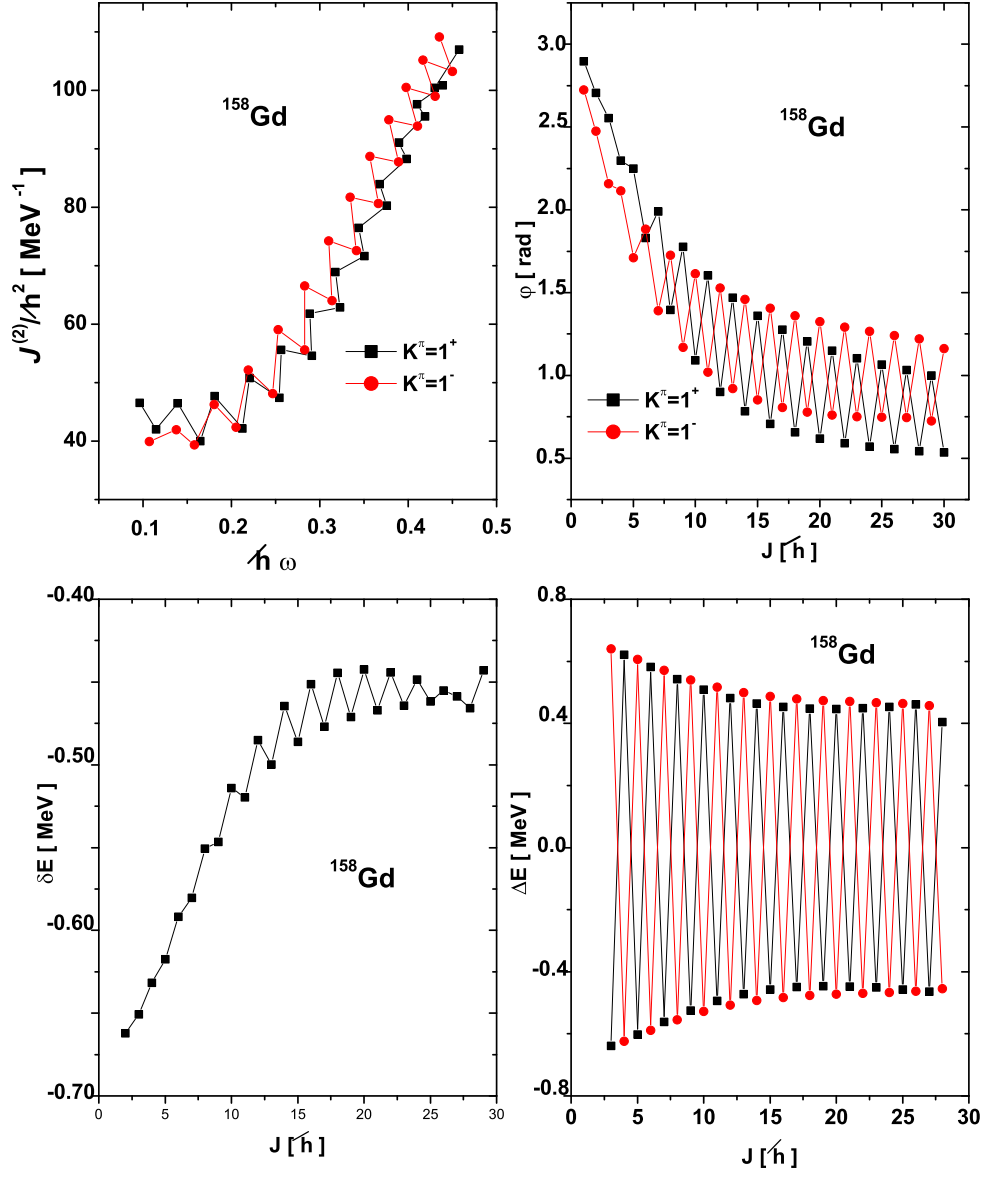}
\end{center}
\caption{ Upper left panel: The dynamic moment of inertia is plotted as function of angular frequency.
Upper right panel: The angle between the angular momenta $\vec{J}_2$ and $\vec{J}_3$
is represented as function of angular momentum. Low-left panel:
the first order energy displacement function
is plotted vs. angular momentum. Low-right panel: the second order energy displacement
is plotted as function of angular momentum. All theoretical results correspond
to $^{158}$Gd. }
\label{Fig. 3}
\end{figure}

\clearpage

\begin{figure}[ht]
\begin{center}
\includegraphics[width=0.8\textwidth]{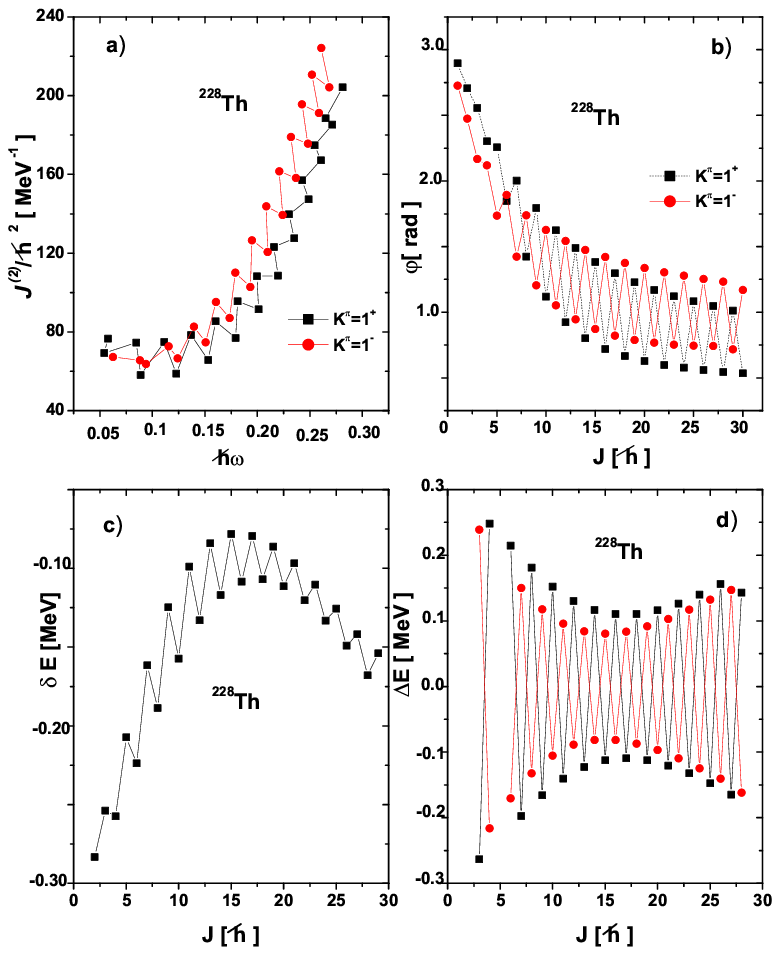}
\end{center}
\caption{The same as in Fig. 3 but for $^{228}$Th.}
\label{Fig. 4}
\end{figure}
\clearpage

\begin{figure}[ht]
\begin{center}
\includegraphics[width=0.8\textwidth]{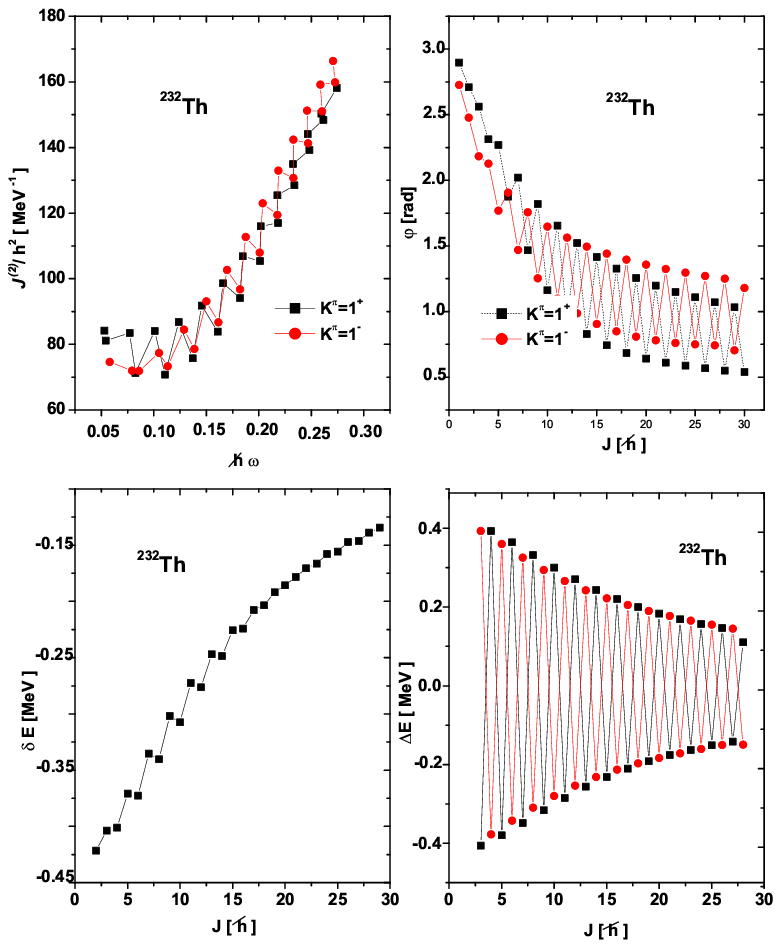}
\end{center}
\caption{The same as in Fig. 3 but for $^{232}$Th.}
\label{Fig. 5}
\end{figure}

\clearpage

\begin{figure}[ht]
\begin{center}
\includegraphics[width=0.8\textwidth]{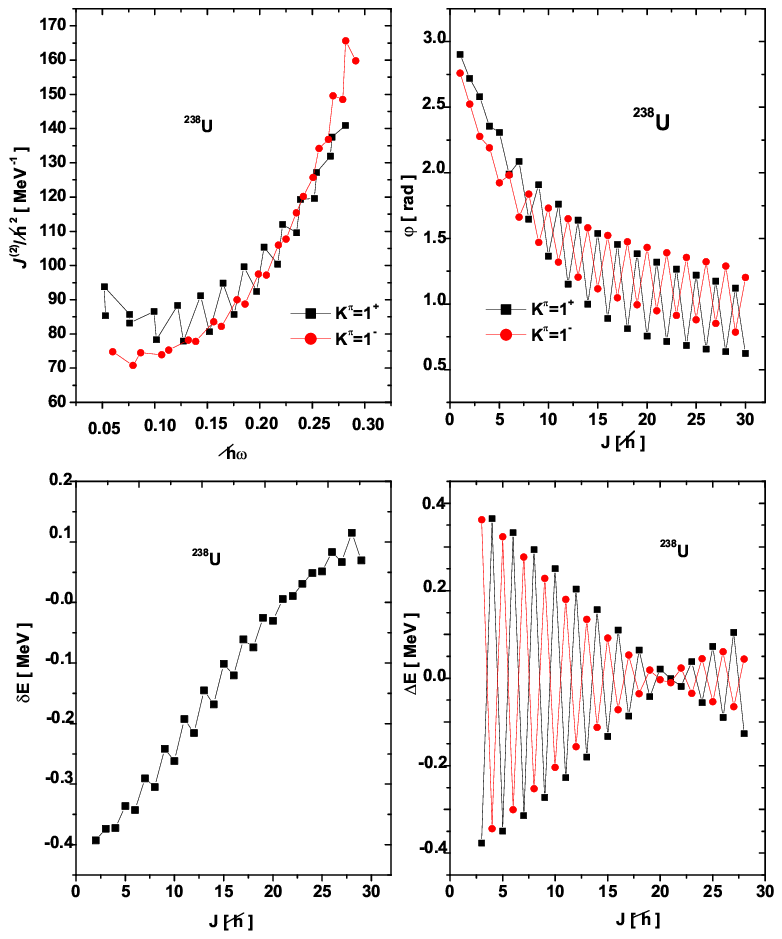}
\end{center}
\caption{The same as in Fig. 3 but for $^{238}$U.}
\label{Fig. 6}
\end{figure}

\clearpage

\begin{figure}[ht]
\begin{center}
\includegraphics[width=0.8\textwidth]{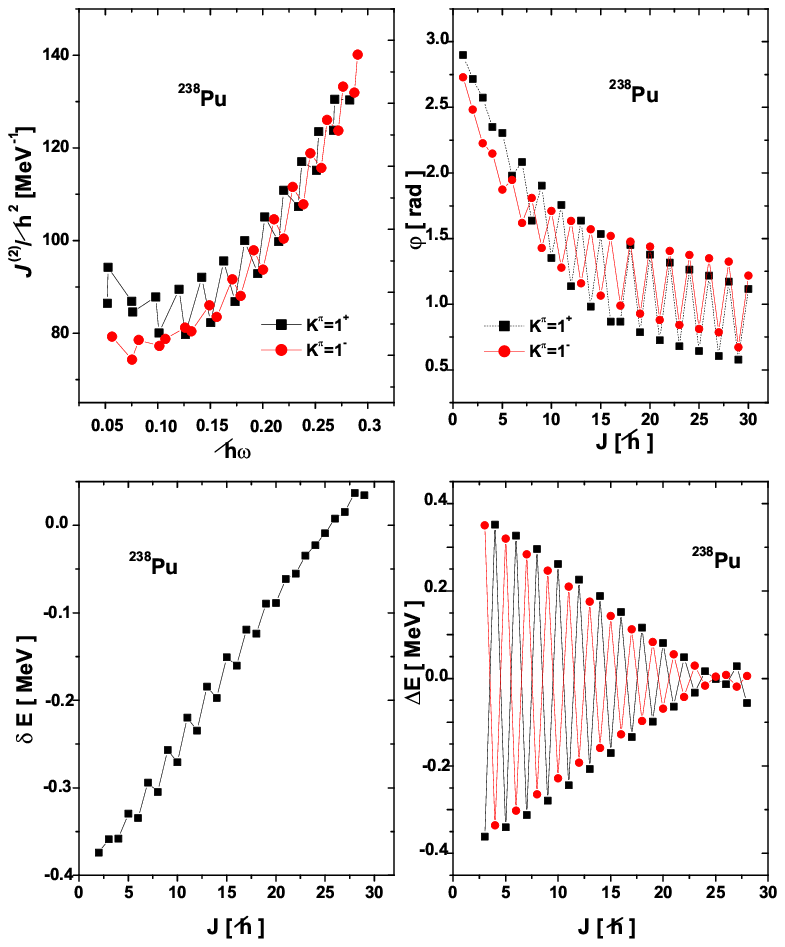}
\end{center}
\caption{The same as in Fig. 3 but for $^{238}$Pu.}
\label{Fig. 7}
\end{figure}

\clearpage

\begin{figure}[ht]
\begin{center}
\includegraphics[width=0.8\textwidth]{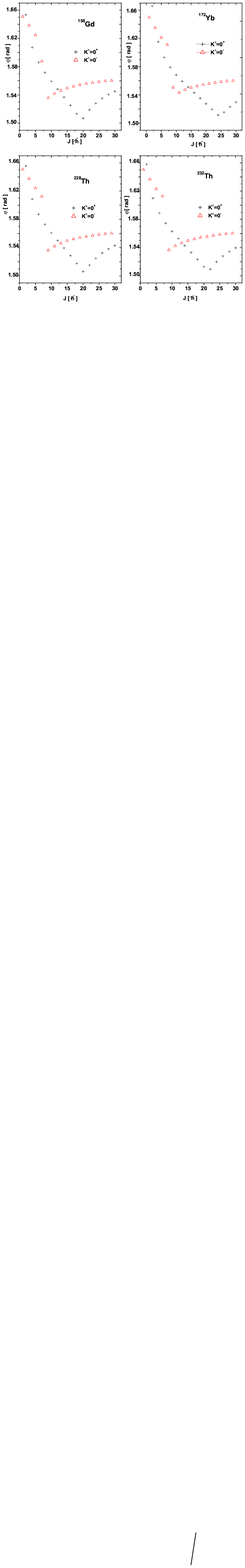}
\end{center}
\caption{\label{Fig. 8}}
\end{figure}

\clearpage
\small{Fig. 8. The angle between quadrupole and octupole angular momenta in the negative (up triangle) and positive (dagger) ground bands for the nuclei mentioned in the four panels}

\clearpage

\begin{figure}[ht]
\begin{center}
\includegraphics[width=0.8\textwidth]{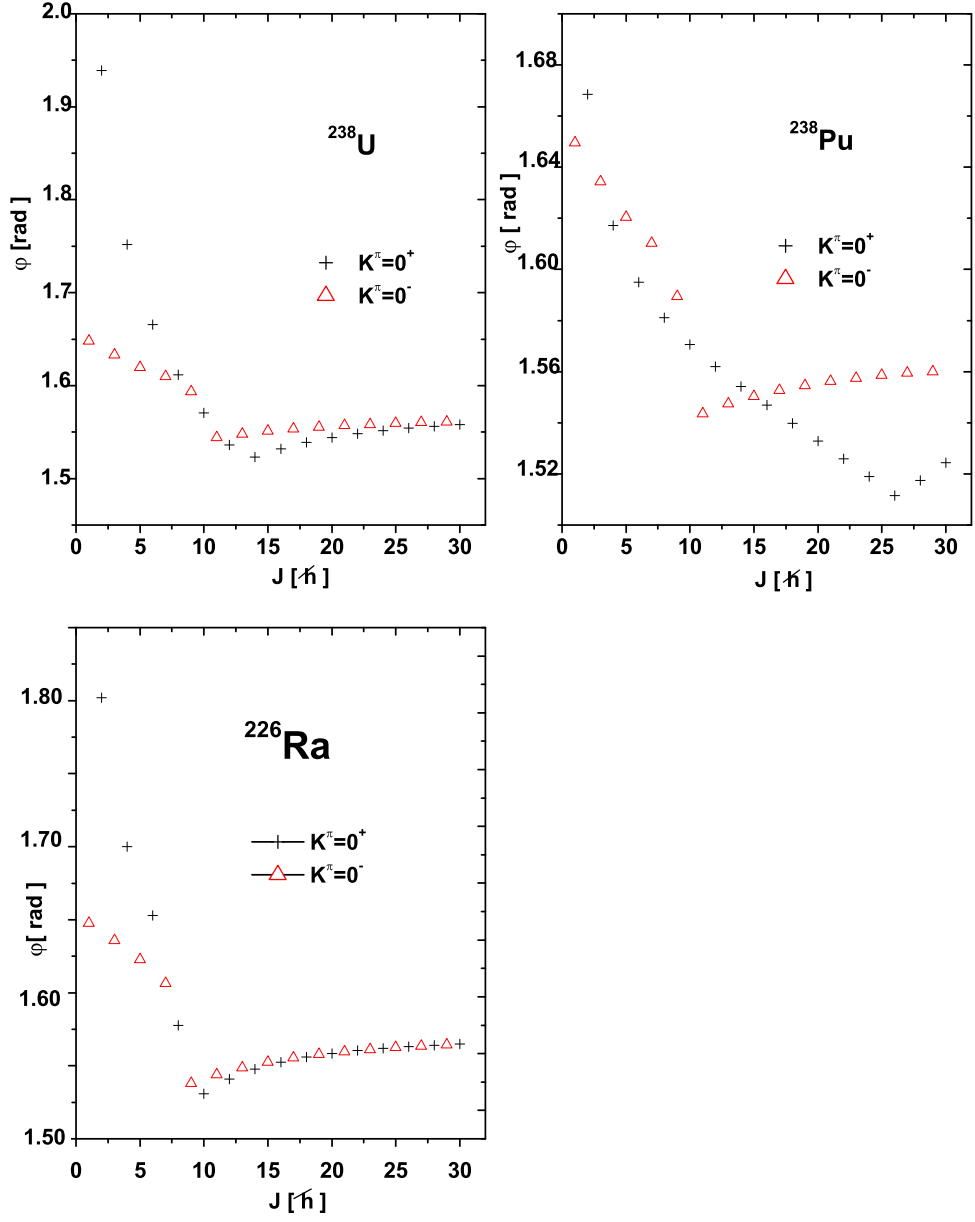}
\end{center}
\caption{The same as in Fig. 8 but for $^{238}$U, $^{238}$Pu and  $^{226}$Ra.}
\label{Fig. 9}
\end{figure}

\begin{figure}[ht]
\begin{center}
\includegraphics[width=0.8\textwidth]{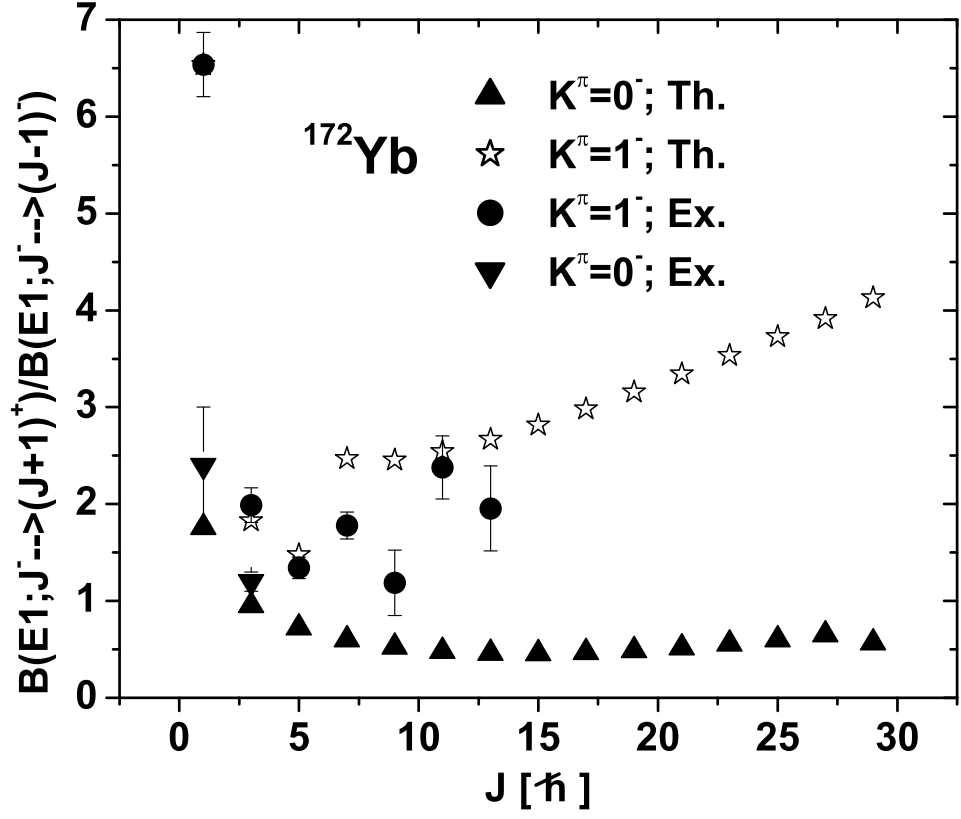}
\end{center}
\caption{The experimental branching ratios for the bands $K^{\pi}=1^-$ (black circle) and
 and $K^{\pi}=0^-$ (down triangle) in $^{172}$Yb are given as function of
 angular momentum.
 For comparison the calculated branching ratios, represented by stars and up triangles
 respectively, are
 given as function of angular momentum. The transition operator involves the anharmonic term given by
 Eq. (4.4)}
\label{Fig. 10}
\end{figure}

\clearpage

\begin{figure}[ht]
\begin{center}
\includegraphics[width=0.8\textwidth]{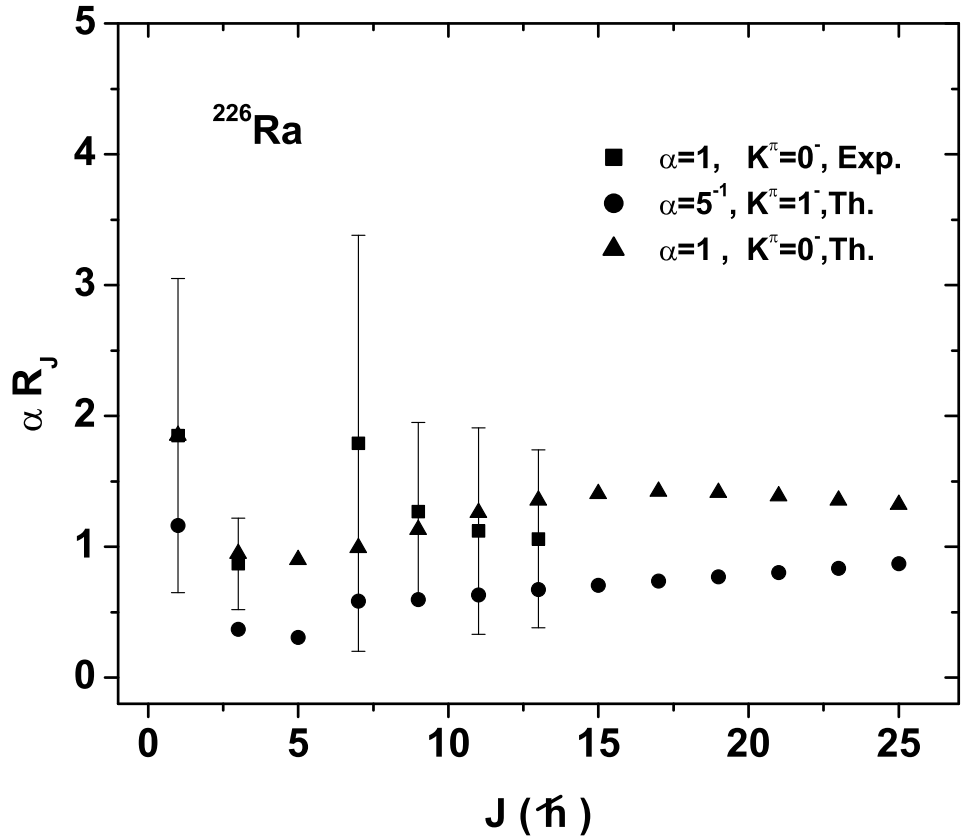}
\end{center}
\caption{ The same as in Fig. 10 but for $^{226}$Ra. For a better
representation the results for the $K^{\pi}=1^-$ band are divided by 5.
The experimental data for this band are lacking. }
\label{Fig. 11}
\end{figure}

\clearpage

\begin{figure}[ht]
\begin{center}
\includegraphics[width=0.8\textwidth]{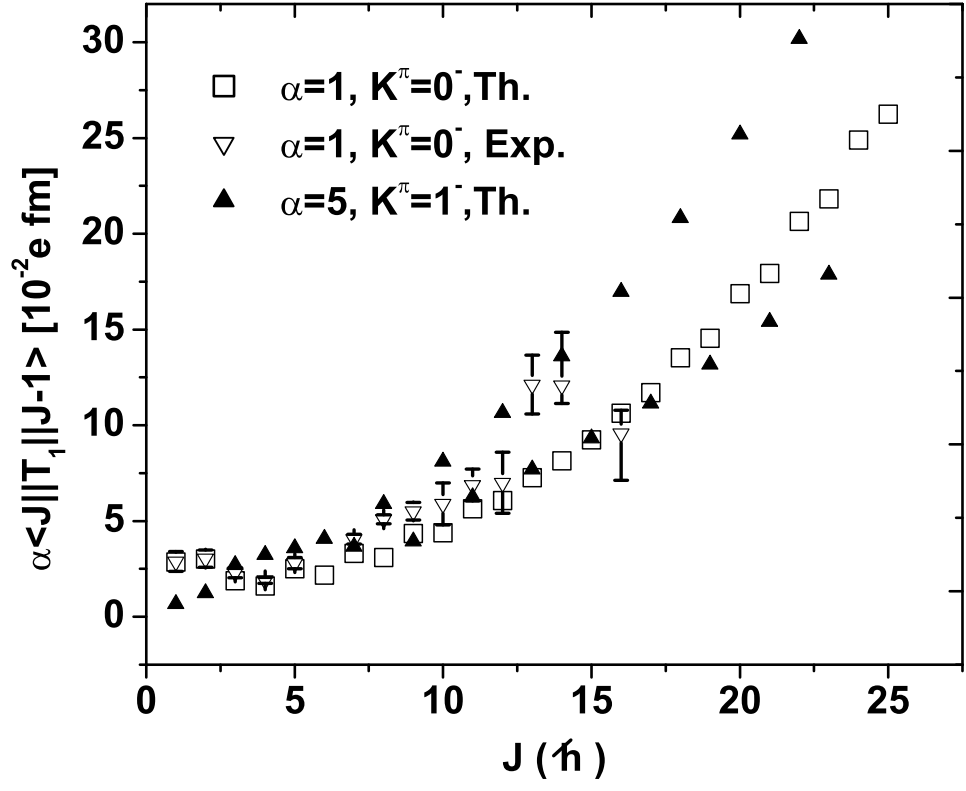}
\end{center}
\caption{ The reduced matrix element for the electric dipole transition
$J^-\to (J-1)^+$ with $J=odd$ and $J^+\to (J-1)^-$ for $J=even$.}
\label{Fig. 12}
\end{figure}

\clearpage

\begin{figure}[h]
\begin{center}
\includegraphics[width=0.8\textwidth]{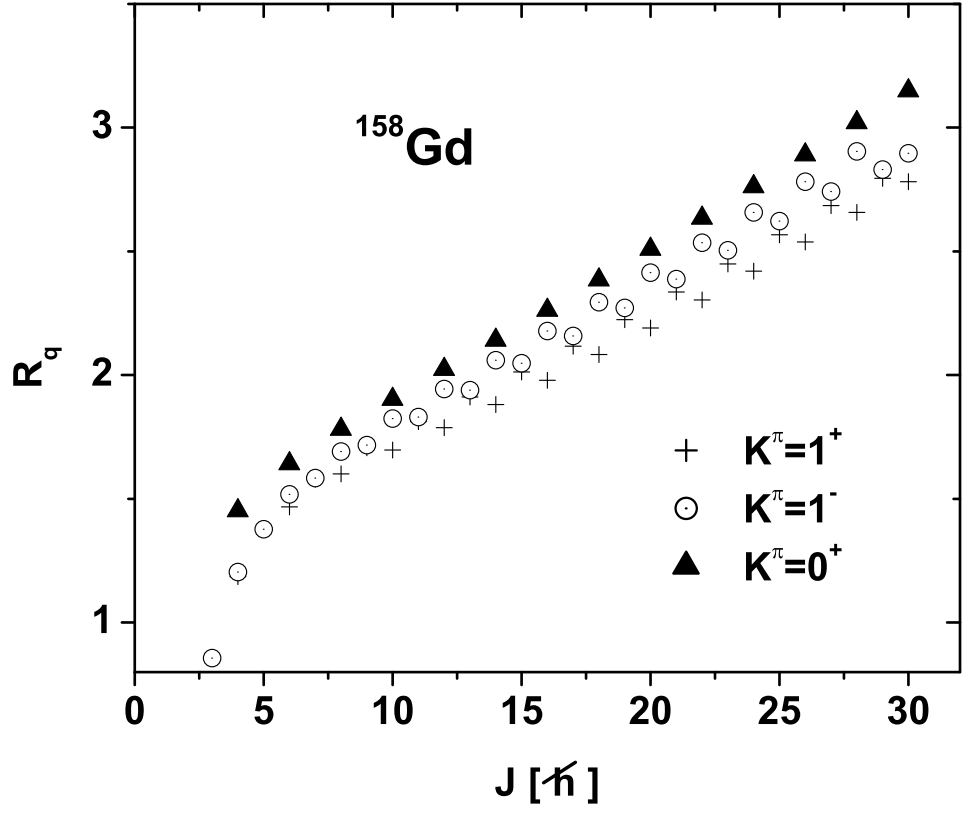}
\end{center}
\caption{The B(E2) values for the intraband transitions for the $K^{\pi}=1^+$ and
 $K^{\pi}=1^-$ divided by the value corresponding to the transition
 $2^{\dagger}_g\to  0^{\dagger}_g$ are given as function of angular momentum.
 The ratio is denoted by $R_q$ according to Eq.(\ref{ErQ}).
 For comparison, the intraband B(E2) values characterising the ground band are also given.}
\label{Fig. 13}
\end{figure}
\clearpage
\end{document}